\documentclass[aps,prd,superscriptaddress,amsmath,groupedaddress,twocolumn,nofootinbib]{revtex4}
\usepackage{color,hhline,psfrag,rotating,amssymb}
\usepackage{dcolumn}

\begin{document}
\title{Nonperturbative comparison of clover and HISQ quarks in lattice QCD and the properties of 
the $\phi$ meson}
\author{Bipasha Chakraborty}
\email[]{bipasha@jlab.org}
\affiliation{SUPA, School of Physics and Astronomy, University of Glasgow, Glasgow, G12 8QQ, UK}
\affiliation{Jefferson Lab, 12000 Jefferson Avenue, Newport News, Virginia 23606, USA}
\author{C. T. H. Davies}
\email[]{christine.davies@glasgow.ac.uk}
\affiliation{SUPA, School of Physics and Astronomy, University of Glasgow, Glasgow, G12 8QQ, UK}
\author{G. C. Donald}
\affiliation{Institute for Theoretical Physics, University of Regensburg, 93040 Regensburg, Germany}
\author{J. Koponen}
\affiliation{SUPA, School of Physics and Astronomy, University of Glasgow, Glasgow, G12 8QQ, UK}
\affiliation{INFN, Sezione di Tor Vergata, Via della Ricerca Scientifica 1, I-00133 Roma, Italy}
\author{G. P. Lepage}
\affiliation{Laboratory of Elementary-Particle Physics, Cornell University, Ithaca, New York 14853, USA}
%\thanks{}
%\affiliation{}
\collaboration{HPQCD collaboration}
\homepage{http://www.physics.gla.ac.uk/HPQCD}
\noaffiliation

\date{\today}

\begin{abstract}
We compare correlators for pseudoscalar and vector mesons made from valence 
strange quarks using the clover quark and 
highly improved staggered quark (HISQ) formalisms in full lattice QCD. 
We use fully nonperturbative methods to normalise vector and axial vector 
current operators made from HISQ quarks, clover quarks and from combining 
HISQ and clover fields. 
This allows us to test expectations for the renormalisation factors 
based on perturbative QCD, with implications for the error budget 
of lattice QCD calculations 
of the matrix elements of clover-staggered $b$-light weak currents, 
as well as further HISQ calculations of the hadronic vacuum polarisation. 
We also compare the approach to the (same) continuum limit in clover and 
HISQ formalisms for the mass and decay constant of the $\phi$ meson. 
Our final results for these parameters, using single-meson correlators 
and allowing an uncertainty for the neglect of quark-line disconnected diagrams 
are: $M_{\phi} =$ 1.023(6) GeV and 
$f_{\phi} = $ 0.238(3) GeV in good agreement with experiment. The 
results come from calculations in the HISQ formalism 
using gluon fields that include the effect of $u$, $d$, $s$ and 
$c$ quarks in the sea with three lattice spacing values and 
$m_{u/d}$ values going down to the physical point. 
 
\end{abstract}

%\pacs{}
%\keywords{}

\maketitle

\section{Introduction}
\label{sec:intro}
Weak decay matrix elements calculated in lattice QCD are critical to the flavour physics programme of overdetermining the Cabibbo-Kobayashi-Maskawa matrix to find signs of new physics (see, for example,~\cite{Davieslat11,Lunghilat11}). 
For this programme it is particularly important to study heavy flavour physics and, although it is now becoming possible to study heavy quarks using relativistic formalisms~\cite{bcHISQ, fBsHISQ}, 
the most extensive studies of heavy quarks in lattice QCD have been done with 
nonrelativistic formalisms (or at least formalisms that make use of nonrelativistic methods), 
such as NRQCD~\cite{nrqcd} or the Fermilab formalism~\cite{fnal}. 
In nonrelativistic formalisms a critical issue is the normalisation of the current operator that couples to the $W$ boson, and this is one of the main sources of error in the lattice QCD result. 
Relativistic formalisms can be chosen to have absolutely normalised currents, for example through the existence of a partially conserved axial current (PCAC) relation~\cite{fdshort}. The main issue with relativistic formalisms is then controlling discretisation errors~\cite{HISQ}.  

The archetypal heavy meson weak decay process is annihilation of a $B$ meson to $\tau \nu$. The hadronic parameter which controls the rate of this process is the $B$ meson decay constant, $f_B$, proportional to the matrix element to create a $B$ meson from the vacuum with the temporal axial current containing a bottom quark field and a light antiquark field. 
When the heavy quark field uses a nonrelativistic formalism the simplest way to match the appropriate current in lattice QCD to that in a continuum scheme is using lattice QCD perturbation theory. 
Such calculations of the $Z$ factors required have been done 
through $\mathcal{O}(\alpha_s)$ for both 
NRQCD~\cite{Znrqcdclover, Znrqcdasqtad, Znrqcdhisq} 
and Fermilab~\cite{Zfnalclover, Zfnalasqtad} heavy quarks with a variety 
of different light quark formalisms. 
The most recent results for $B$ meson decay constants using NRQCD are given in~\cite{Dowdall:2013tga} 
and using Fermilab heavy quarks in~\cite{Bazavov:2011aa}. 

In doing these calculations for Fermilab heavy quarks and clover light quarks~\cite{Zfnalclover} it 
was noticed that the heavy-light current renormalisation differed very little at 
$\mathcal{O}(\alpha_s)$ from the square root of the product of $Z$ factors for the 
temporal vector heavy-heavy and light-light currents, which can be determined nonperturbatively. 
This then gives rise to the possibility of determining, for example, $Z_{{A^4}_{hl}}$ with small uncertainty 
if it can be demonstrated that this result is true to all orders in perturbation theory and 
is not specific to only one light quark formalism (or heavy quark formalism). 
This question is a critical one for 
the reliability of the estimates of perturbative errors in determinations 
of $f_B$ and $f_{B_s}$ and other weak matrix elements using this approach.    
The same issues arise, for example, for the vector current with implications 
for the matrix elements calculated for $B\rightarrow \pi \ell \nu$ from 
lattice QCD~\cite{Lattice:2015tia}. 

Here we test this fully nonperturbatively for the case where the `heavy-light' current is 
made of a clover quark ($\equiv$ Fermilab formalism at low mass) and a highly improved 
staggered quark (HISQ)~\cite{HISQ} 
both tuned accurately to the strange quark mass, following the suggestion in~\cite{Davieslat11}. 
We use the absolute normalisation for the HISQ-HISQ temporal axial vector current 
that arises from chiral symmetry in that formalism 
to normalise both the HISQ-clover and clover-clover temporal axial 
vector current. By determining the normalisation of the appropriate
vector currents, also 
fully nonperturbatively, we can then determine the ratio used by the Fermilab collaboration and 
test it against the hypothesis that it should be close to 1. 

From the same $s$ quark propagators for the study above we can also make vector ($\phi$) meson correlators and study the $\phi$ meson mass and decay constant for the cases where the $\phi$ is made purely of 
clover quarks or purely of HISQ quarks, or made of one of each. Our results cover 
3 values of the lattice spacing spanning the range from 0.15 fm to 0.09 fm and so 
we can compare the approach to the continuum limit of the two formalisms (and test whether 
they have a common continuum limit) for the 
two calculations. 

Finally we make a more extensive analysis of the $\phi$ meson using the HISQ formalism covering a more 
complete range of gluon field ensembles that includes multiple values of the $u/d$ quark 
mass in the sea going down to the physical value, and allowing physical results 
to be derived. Our calculation uses single-meson correlators only 
and neglects quark-line disconnected diagrams (which we expect 
to have negligible impact).  
Our results tend to confirm that 
the impact of coupling the $\phi$ to its $K\overline{K}$ decay mode is small and 
increases the $u/d$ quark mass-dependence of the $\phi$ properties determined in 
lattice QCD. We are able to 
obtain the $\phi$ mass and decay constant to an accuracy of a few MeV and 
in agreement with experiment. 
Understanding the properties of the $\phi$ from lattice QCD is important 
because it provides a good vector final state for alternative 
studies of semileptonic 
weak decay rates compared to the usual pseudoscalar final states. 
For example, $V_{cs}$ can be determined from $D_s \rightarrow \phi \ell \nu$ 
given lattice QCD results and experimental rates~\cite{babar, Donald:2013pea, Hietala:2015jqa}. 
$B_s \rightarrow \phi \ell^+ \ell^-$ is potentially an important rare decay mode for 
searches for new physics~\cite{Horgan:2013pva}.  

The paper is laid out as follows: Section~\ref{sec:back} describes the background to our calculation; 
the perturbative studies of the renormalisation factors that have been done for current operators 
using different actions and combinations of actions, and the general picture that emerges 
that needs to be tested nonperturbatively. 
Section~\ref{sec:latt} describes our lattice calculation to do these tests and gives our results 
for the nonperturbative determination of $Z$ factors for the HISQ-clover and clover-clover case, 
showing how the nonperturbative determination backs up the picture seen perturbatively. 
We also compare discretisation effects in the clover and HISQ formalisms through the properties 
of the $\phi$ meson using 
the Z factors we have obtained to normalise the decay constant. 
Section~\ref{sec:phi} gives our results for the mass and decay constant of the $\phi$ in the 
HISQ formalism only, covering 
$u/d$ quark masses down to the physical value and allowing a chiral/continuum extrapolation 
to the physical point. 
Section~\ref{sec:conclusions} gives our conclusions. 
Appendix~\ref{appendix:nrqcd} considers the renormalisation factors for currents 
with NRQCD heavy quarks and Appendix~\ref{subsec:Zsf} uses our results for the renormalisation factors for local vector currents for HISQ quarks to extrapolate to values on finer lattices. 

\section{Background}
\label{sec:back}

To provide accurate physical results for hadronic matrix elements, 
lattice QCD current operators must be renormalised to match to those in continuum 
QCD. For some currents and quark formalisms absolute normalisation is 
possible; for example for the temporal axial current in formalisms 
with sufficient chiral symmetry. In other cases a renormalisation 
$Z$ factor must be determined as accurately as possible. Since the 
$Z$ factor, beyond tree-level, allows for the difference between 
gluon radiation in the continuum 
and that in the presence of the lattice momentum cut-off, it is   
an ultra-violet quantity and can be determined in QCD perturbation 
theory. Lattice QCD perturbation theory is relatively complicated 
and such calculations have generally been restricted to the determination 
of effects at $\mathcal{O}(\alpha_s)$ only. 
$Z$ is then determined by equating the one-loop 
scattering amplitude between on-shell quark states in continuum 
QCD and on the lattice. 

Early calculations in which a heavy quark in the Fermilab formalism~\cite{fnal} was
combined with a clover light quark 
found that the heavy-light current renormalisation~\cite{Zfnalclover} differed very little at 
$\mathcal{O}(\alpha_s)$ from the square root of the product of $Z$ factors for the 
temporal vector heavy-heavy and light-light currents. 
This was found also to be true for Fermilab heavy quarks and asqtad 
light quarks~\cite{Zfnalasqtad}. Specifically, the Fermilab Lattice collaboration writes 
for the temporal axial vector current:
\begin{equation}
Z_{{A^4}_{hl}} = \rho \sqrt{Z_{{V^4}_{hh}}Z_{{V^4}_{ll}}}
\label{eq:zhl}
\end{equation}
where 
\begin{equation}
\rho = 1 + \rho^{(1)}\alpha_s + \rho^{(2)}\alpha_s^2 + \ldots 
\label{eq:rho}
\end{equation}
and $\rho^{(1)}$ is found to be very small (typically $< 4\pi\times 0.01$) if the heavy quark mass is not too large. Note then that this is a relationship valid for `light' heavy 
quarks and not in the infinite quark mass (static) limit. 
In practice the region of small values of $\rho^{(1)}$ extends for heavy 
quark masses, $m_h$, in the Fermilab formalism up to the $b$ quark mass at 
least for fine lattices, with $a < 0.1\,\mathrm{fm}$. For small values of 
$m_h$ the Fermilab formalism becomes identical to the standard tadpole-improved clover formalism.

$Z_{{V^4}_{hh}}$ and $Z_{{V^4}_{ll}}$ are the renormalisation factors for local temporal vector 
currents made respectively of Fermilab formalism quarks and light quarks in whatever 
formalism is being used for the heavy-light current. 
These vector current $Z$ factors can be determined fully nonperturbatively in lattice QCD by 
demanding normalisation of the vector form factor between two identical mesons at rest. 

Eq.~(\ref{eq:zhl}) then gives rise to the possibility that $Z_{{A^4}_{hl}}$ can be 
determined with small errors if it can be shown that $\rho$ is indeed close to 1 to 
all orders in perturbation theory. The argument that this should be true is based on 
the idea that a large part of the perturbative $Z$ comes from the self-energy of the 
individual quark legs and this part will cancel in $\rho$~\cite{Zfnalclover}. This 
cancellation will include tree-level mass dependence and tadpole effects. 
However, this only guarantees that $\rho^{(2)}$ and higher coefficients should be 
`of reasonable size', not that they should be as small as $\rho^{(1)}$ is found to be. 
The question of what uncertainty it is reasonable to take for 
the missing $\alpha_s^2$ and higher order pieces 
is then a critical one for the reliability of the estimates 
of perturbative errors in determinations of $f_B$ and $f_{B_s}$ and other 
weak matrix elements using this approach.    

In testing this relationship nonperturbatively we note 
that to be robust it must be fairly general and work 
for a variety of formalisms, for example any light quark 
formalism combined with a Fermilab formalism heavy quark. 
Since in fact it is a relationship that works best for 
light quarks in the Fermilab formalism, we can 
substitute standard clover quarks for Fermilab 
quarks since the Fermilab formalism becomes the 
clover formalism in the light quark mass limit. 
This avoids then any need to handle $\Lambda/m_h$ (where 
$m_h$ is the heavy quark mass) corrections to 
the `heavy-light' currents. 

We then test eq.~(\ref{eq:zhl}) for the case where the 
current on the lefthandside contains two light quarks 
that use different formalisms. One formalism is clover, 
representing the Fermilab formalism. For the other formalism
we could use the asqtad staggered formalism to test 
directly the results from~\cite{Zfnalasqtad}. However it 
makes more sense to use the current state-of-the-art staggered  
formalism, HISQ~\cite{HISQ}, since we will also use the 
state-of-the-art MILC collaboration gluon field 
configurations that include $u$, $d$, $s$ and $c$ 
quarks in the sea using the HISQ formalism. 
We will tune the masses of the valence light quarks to that 
of the strange quark because this can be done very 
accurately~\cite{Davies:2009tsa, Dowdall:2013rya} using 
the pseudoscalar `strange-onium' meson, the $\eta_s$ and will 
give higher 
statistical accuracy for this test than using lighter quarks. 

Because the HISQ formalism has a remnant chiral symmetry 
it has an absolutely normalised temporal axial current. 
By comparing the matrix element between the vacuum and 
the $\eta_s$ of temporal axial currents made of clover 
quarks or mixed currents with one clover and one HISQ 
quark to that made of HISQ quarks we can determine the 
$Z$ factor for the clover-clover current and the HISQ-clover 
current. We can also readily determine the $Z$ factors 
for the local temporal vector current made of HISQ quarks 
or of clover quarks, or the mixed HISQ-clover current, 
by setting the vector form factor 
to 1 between two $\eta_s$ mesons made of appropriate quark 
formalisms at rest.  

We then have all the $Z$ factors necessary to test 
the relationship equivalent to eq.~(\ref{eq:zhl}):
\begin{equation}
Z_{{J}_{\mathrm{H-cl}}} = \rho_J \sqrt{Z_{{V^4}_{\mathrm{cl-cl}}}Z_{{V^4}_{\mathrm{H-H}}}},
\label{eq:zJ}
\end{equation}
where H stands for HISQ and cl for clover, 
for the cases where the current $J$ is the temporal 
axial current or the temporal vector current. 
In both cases we can determine how close to 1 
$\rho_J$ is and therefore how small the perturbative 
coefficients that make up $\rho_J$ must be. 

As a side-product of these calculations we can 
test a number of other relationships between 
$Z$ factors, including that between the temporal 
axial vector and temporal vector currents in all 
three combinations of formalisms, H-H, H-cl and cl-cl. 
Note that the $Z$ factor being determined on the lefthandside 
of eq.~(\ref{eq:zhl}) is a flavour-nonsinglet current. 
Our equivalent expression, implied by eq.~(\ref{eq:zJ}), then also 
corresponds to a flavour-nonsinglet current even though both 
quarks are $s$ quarks. This means that we do not need to consider 
any quark-line disconnected contributions to the correlation functions 
that we are using for this analysis. The $Z_V$ factors on the 
righthandside of eq.~(\ref{eq:zhl}) correspond to vector currents 
for quarks of the same flavour; in this case quark-line disconnected 
contributions are negligible~\cite{Bakeyev:2003ff} and can be ignored.

The next section describes the lattice calculation 
and gives results for these $Z$ factors. 

\section{Z factors}
\label{sec:latt}

\subsection{Lattice configurations and simulation parameters}
\label{subsec:par}
\begin{table}  
\begin{tabular}{llllllll}
\hline
\hline
Set&$\beta$&$w_0/a$&$am_{l}^{sea}$&$am_{s}^{sea}$&$am_c^{sea}$&$L_s/a$&$L_t/a$\\
\hline
1 & 5.80 & 1.1119(10) & 0.013 &0.065 & 0.838 & 16 & 48\\ 
2 & 5.80 & 1.1367(5) & 0.00235 &0.0647 & 0.831 & 16 & 48\\ 
\hline
3 & 6.00 & 1.3826(11) & 0.0102 & 0.0509 & 0.635 & 24 & 64\\ 
4 & 6.00 & 1.4029(9) & 0.00507 & 0.0507 & 0.628 & 24 & 64\\ 
5 & 6.00 & 1.4029(9) & 0.00507 & 0.0507 & 0.628 & 32 & 64\\ 
6 & 6.00 & 1.4029(9) & 0.00507 & 0.0507 & 0.628 & 40 & 64\\ 
7 & 6.00 & 1.4149(6) & 0.00184 & 0.0507 & 0.628 & 48 & 64\\ 
\hline
8 & 6.30 & 1.9006(20) & 0.0074 & 0.0370 & 0.440 & 32 & 96\\ 
9 & 6.30 & 1.9518(7) & 0.0012 & 0.0363 & 0.432 & 64 & 96\\ 
\hline
\hline
\end{tabular}
\caption{Sets of MILC configurations used here with their (HISQ) sea quark masses, 
$m_l$ ($m_u=m_d=m_l$), $m_s$ and $m_c$ in lattice units. $\beta=10/g^2$ is the QCD gauge 
coupling and $w_0/a$~\cite{Dowdall:2013rya, Chakraborty:2014aca} gives the lattice spacing, $a$, in terms of 
the Wilson flow parameter, $w_0$~\cite{Borsanyi:2012zs}.
The lattice spacing is approximately 
0.15 fm for sets 1 and 2; 0.12 fm for sets 3-7 and 0.09 fm for sets 8 and 9. 
The lattice size is $L_s^3 \times L_t$. Ensemble sizes are 500 to 1000 
configurations each.}
\label{tab:params}
\end{table}

We use gluon field ensembles generated by the MILC collaboration~\cite{Bazavov:2012uw} at widely 
differing values of the lattice spacings: 0.15 fm, 0.12 fm and 0.09 fm. 
The relative lattice spacings were fixed using a determination of $w_0/a$~\cite{Dowdall:2013rya}
\footnote{Note that the value 
on set 8 has changed from that given in~\cite{Dowdall:2013rya}; we are grateful 
to C. McNeile for providing this updated value.} 
 where $w_0$ 
is the Wilson flow parameter~\cite{Borsanyi:2012zs}. The absolute value of $w_0$ 
was determined from $f_{\pi}$~\cite{Dowdall:2013rya} to be 0.1715(9) fm. 
The gluon field ensembles include the effect of $u$, $d$, $s$ and $c$ quarks in 
the sea (with degenerate $u$ and $d$ quarks) 
using the HISQ formalism and also use a gluon action improved fully 
through $\mathcal{O}(\alpha_sa^2)$~\cite{Hart:2008sq}. 
We therefore expect the gluon fields to have very small `intrinsic' 
discretisation errors which is useful for studying the discretisation 
errors of meson correlation functions made on these configurations using different quark formalisms. 

For our determination of clover $Z$ factors we have chosen to use the ensembles 
1, 3 and 8 that have a sea light quark mass in units of the sea strange mass $m_l/m_s = 0.2$. 
This is for reasons of numerical speed since these lattices have relatively modest
size of 3.5 fm. Since we are calculating meson correlation functions made purely 
of strange quarks, we expect sea quark mass effects to be small so the fact that 
$m_l^{\mathrm{sea}}$ is not physical is not an issue. Finite volume effects were
shown to be negligible for the $\eta_s$ for lattices of size 3.5 fm 
in~\cite{Chakraborty:2014mwa} (see also Section~\ref{sec:phi}).
In any case we would expect such effects to be the same for the HISQ and clover valence 
quarks and hence any effects should cancel in the ratios we use to determine 
$Z$ factors. 

On gluon field ensembles 1, 3 and 8 we calculate valence HISQ and clover 
quark propagators using the standard HISQ action~\cite{HISQ} (as used for the 
sea quarks) and the standard tadpole-improved space-and-time-symmetric 
clover action used for light quarks~\cite{DeGrand:2006zz}. 
In the clover action the gluon fields $U_{\mu}$ are divided by 
a tadpole parameter~\cite{Lepage:1992xa}, $u_0$, for which we use the fourth root of the plaquette.
The parameter values are listed in Table~\ref{tab:valence}. 

\begin{table}
\begin{tabular}{lllllll}
\hline
\hline
Set  & ${am_s}^{H,val}$ & ${\kappa_s}^{cl,val}$ & $u_0$ & $n_{cfg}$ & $n_t$ & 3pt T  \\
\hline
1 &  0.0705 & 0.14082 & 0.85535 & 1021 & 12 & 9, 12, 15, 18\\
\hline
3 &  0.0541 & 0.13990 & 0.86372 & 527 & 16 & 12, 15, 18, 21\\
\hline
8 &  0.0376 & 0.13862 & 0.87417 & 504 & 16 & 16, 19, 22, 25\\
\hline
\hline
\end{tabular}
\caption{List of parameters used for the valence quarks. Column 2 gives the HISQ bare mass. 
Columns 3 and 4 give the clover $\kappa$ value and the tadpole factor $u_0$ used to 
tadpole-improve the action. Column 5 gives the number of configurations used for 
most of the calculations and 
column 6 the number of time sources on each configuration. Because our 
HISQ valence quarks are much faster to calculate we have determined $\eta_s$ H-H 
correlators on double the number of configurations for sets 3 and 8. We only 
determined the 3-point correlators for the H-cl current on half of 
the configurations on set 8, however. The 
final column gives the T values used in the determination of 3-point correlation 
functions. }
\label{tab:valence}
\end{table}

For the source for each propagator we divide the spatial slice of the lattice at a given 
time value into $2^3$ cubes and use a Gaussian random number for each color at the spatial points 
corresponding to the corners of each cube. 
We use many time sources on each 
configuration to improve statistics (see Table~\ref{tab:valence}) and they are evenly spaced 
through the lattice. The starting time source for each configuration is chosen randomly to 
reduce autocorrrelations, which are small for $\eta_s$ correlators~\cite{Dowdall:2011wh}. 

\begin{figure}
\centering
\includegraphics[width=0.45\textwidth]{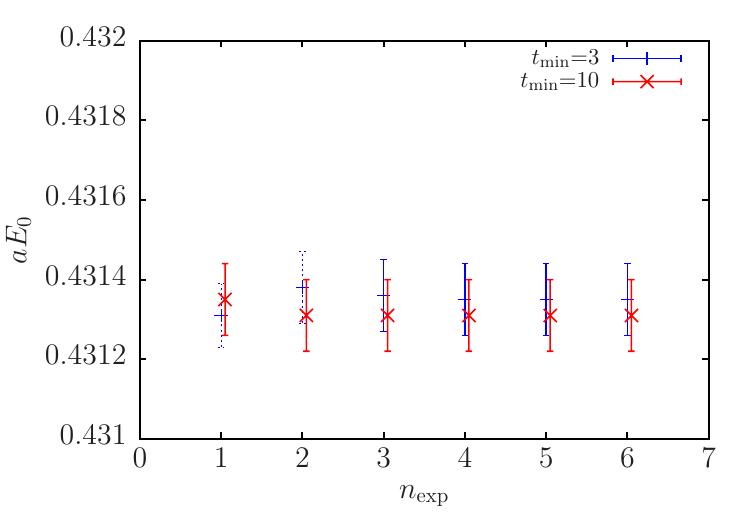}
\caption{Results for the ground-state energy, $E_0$ in lattice units, 
for the H-H $\eta_s$ on coarse set 3 as a function of the number of exponentials 
used in the fit (eq.~(\ref{eq:hisqetafit})). 
We show results for a $t_{\mathrm{min}}$ value of 3 and 10; 
the results are shown with dashed lines for fits where the 
$\chi^2/[\mathrm{dof}]>1$. 
}
\label{fig:massfit}
\end{figure}

We combine the 
HISQ propagator with its complex conjugate into a pseudoscalar 
meson correlator (two-point function) that corresponds 
to the `Goldstone taste' in the parlance of staggered quarks. In spin-taste 
notation this is $\gamma_5 \otimes \gamma_5$ and the correlator simply 
corresponds to the modulus squared of the propagator, summed over a spatial 
slice of the lattice to project onto zero spatial momentum. 
We will denote the ground-state particle of this correlator $\eta_s^{\mathrm{H-H}}$. 
To obtain the ground-state parameters we fit the correlator to the 
standard multi-exponential form as a function of time separation $t$ 
between the source and sink: 
\begin{eqnarray}
\label{eq:hisqetafit}
C_{2pt}&=&\sum_{k=0}^{n_{\mathrm{exp}}-1} a_k^2f(E_k,t); \nonumber \\
f(E_k,t)&=&e^{-E_kt}+e^{-E_k(L_t-t)}. 
\end{eqnarray}
There are no staggered quark `oscillating' terms in the $\eta_s$ correlator 
because it is of Goldstone taste and made of equal mass quarks. 
Our fits use Bayesian methods~\cite{gplbayes} that allow us to include multiple 
exponentials and consequently allow for systematic errors in 
our ground-state parameters from contamination from excited states. 
We use a prior width on all of the amplitudes of 0.5 (larger 
than any of our ground-state amplitudes)
and on the ground-state energy of 0.05 (much larger than any of 
our fit uncertainties on this parameter). 
On the 
energy differences between consecutive states we take a prior 
of 0.8(0.4) GeV (converted back 
to lattice units in the fit). 
We have checked that the ground-state parameters from our fit are 
very insensitive to the priors.  
We drop the very small $t$ values from 
the fit, taking $t_{\mathrm{min}}$ of 3 or 4.
Fit results and uncertainties are stable from 3 or 4 
exponentials upwards with $\chi^2/{\mathrm{dof}}$ varying from 0.5 to 0.9. 
We take our final values from the 6 exponential fit. 
Neither the number of exponentials in the fit, nor the $t_{\mathrm{min}}$ 
value have any significant effect on the result for ground-state 
parameters. We illustrate this in Figure~\ref{fig:massfit}, giving 
the ground-state energy from the fit as a function of the number of exponentials 
included for both $t_{\mathrm{min}}$ of 3 (the value we used) and 
$t_{\mathrm{min}}=10$. For $t_{\min}$ of 3 fits with a small number 
of exponentials (1 and 2) give a poor fit because higher states 
contribute to the correlator at small $t$ values. However, once the 
fit does have a good $\chi^2$ it remains stable as further states 
are added to the fit. For $t_{\mathrm{min}}$ of 10 a good fit can be obtained 
with fewer states included and it agrees with the result using $t_{\mathrm{min}}=3$. 
We prefer to take the smaller $t_{\mathrm{min}}$ value for uniformity 
of fits across all the 2- and 3-point functions we study here. 

Here we are concerned with the properties of the ground-state, 
which  are given by $k=0$. These are the mass of the $\eta_s^{\mathrm{H-H}}$ which 
is given in lattice units by $E_0$ and its decay constant which is determined 
from the ground-state amplitude, $a_0$, as described in section~\ref{subsec:A0}. 

Earlier results~\cite{Dowdall:2013rya} using a variety of both $u/d$ and $s$ HISQ 
valence masses on the more complete set of ensembles from Table~\ref{tab:params} 
allowed us to determine the value of the $\eta_s$ mass in the continuum and 
chiral limits of full lattice QCD. Although the $\eta_s$ meson is not a physical 
particle (because we do not allow it to mix with other flavourless pseudoscalars) 
it is nevertheless useful in lattice QCD for tuning the $s$ quark mass~\cite{Davies:2009tsa}. 
In~\cite{Dowdall:2013rya} we obtained a physical value for the $\eta_s$ mass of 
688.5(2.2) MeV. Here we then tune the bare quark mass in our HISQ action to 
obtain this value for the $\eta_s^{\mathrm{H-H}}$ mass (combining our results 
for $E_0$ from eq.~(\ref{eq:hisqetafit}) with the values of the lattice spacing 
from Table~\ref{tab:params}) on each ensemble. The bare valence quark masses 
obtained are given in Table~\ref{tab:valence}. Note that these values are not 
the same as those used in~\cite{Dowdall:2013rya} because, with the benefit 
of those results, we have improved the tuning (see also~\cite{Chakraborty:2014aca}).   
The $\eta_s$ mass values in lattice units ($E_0$ from our fits) are given 
in Table~\ref{tab:etas}. The precision of the values shows how well 
this tuning can be done. 

We also combine clover quark propagators with their complex conjugates to 
make $\eta_s$ correlators using either the temporal
axial current, $\overline{\psi}\gamma_4\gamma_5\psi$, or 
the pseudoscalar current, $\overline{\psi}\gamma_5\psi$, at both source and 
sink. 
We then fit these correlators simultaneously to the same fit form, eq.~(\ref{eq:hisqetafit}),
given earlier for the H-H case and using the same priors. 
We require both correlators to have the same energies but different 
amplitudes, $a_{k,A^4}$ and $a_{k,P}$.
Again the ground-state parameters are given by $k=0$ and are the ones 
we use here. The ground-state $\eta_s^{\mathrm{cl-cl}}$ mass is given by 
combining values for $E_0$ with the inverse lattice spacing obtained 
from Table~\ref{tab:params}. The mass parameter in the clover action 
is denoted by $\kappa$ with the bare quark mass being related to 
$1/(2\kappa)$ by an additive constant~\cite{DeGrand:2006zz}. 
We tune $\kappa$ to give the same $\eta_s$ 
mass as that discussed for the H-H case above. 
Table~\ref{tab:valence} gives the tuned $\kappa$ values we obtain and 
Table~\ref{tab:etas} gives the $\eta_s$ masses in lattice units ($E_0$ 
from our fits). Again we are able to perform this tuning very precisely. 

The third option is to combine a clover and HISQ propagator 
to make a mixed-action correlator. To do this the HISQ propagators, 
which have no spin component, must be converted back to naive 
quark propagators with a spin component by `undoing' the staggering 
transformation used to obtain the staggered quark action~\cite{Wingate:2002fh, HISQ}. 
Because we have used a `corner wall' source for 
our propagators, with one point per $2^3$ block, the matrices implementing the 
staggered transformation at the source are all the 
unit matrix, which simplifies the combination. 
Once converted to a naive form with 4 spin components the HISQ 
propagators can be straightforwardly combined with clover propagators 
as in the clover-clover case above and using a temporal axial current 
operator at source and sink, or a pseudoscalar operator. 
To fit these correlators (simultaneously) we must include oscillating 
terms that arise from the staggered quark formalism. 
The fit form is then
\begin{eqnarray}
\label{eq:Hclsymmfit}
C_{2pt}(t) &=& \sum_{k=0}^{n_{exp}}a_k^2f(E_k,t) \\
&& -(-1)^{t/a} \sum_{ko=0}^{n_{exp}}a_{ko}^2f(E_{ko},t) \nonumber 
\end{eqnarray}
with normal (non-oscillating) amplitude parameters $a_k$, and amplitudes for oscillating 
terms given by $a_{ko}$.
Again we use priors for the normal terms that are the 
same as those given above for both the H-H and cl-cl cases. For the oscillating 
terms we use the same amplitude and energy difference priors as for the 
normal terms and we take the difference between the energy for the ground-state 
in the oscillating channel and that in the normal channel to be 
0.6(4) GeV.  
We again take the fit 
results from the 6 exponential fit, given stability of the results 
from the 3 or 4 exponential fit upwards. Since the mass parameters 
have now all been tuned, the mass we obtain for the ground-state 
particle in this H-cl channel gives us information about 
discretisation effects. These masses are given in Table~\ref{tab:etas} 
and we can see that they become increasingly close to the 
masses for the H-H and cl-cl channels as the lattice spacing 
becomes smaller moving from set 1 to set 8. This will be 
discussed further in Section~\ref{subsec:disc}. 

\subsection{Z factors for $A^4$}
\label{subsec:A0}

\begin{table}
\begin{tabular}{llllll}
\hline
\hline
Set & Action &  $aM_{\eta_s}$  &  $af_{\eta_s}$ & $af_{\eta_s}/Z_{A^4}$ & $Z_{A^4}$ \\
& comb'n & & & & \\
\hline
1 & H-H & 0.54024(15) & 0.14259(8) &  - \\
  & cl-cl & 0.53966(30) & & 0.19682(26) & 0.7245(10) \\
  & H-cl & 0.57330(24) & & 0.16303(24) & 0.8746(13) \\
\hline 
3 & H-H & 0.43135(9) & 0.11399(4) & - \\
  & cl-cl & 0.43141(20) & & 0.15242(18) & 0.7478(9) \\
  & H-cl & 0.44698(17) & & 0.12946(16) & 0.8804(11) \\
\hline
8 & H-H & 0.31389(7) & 0.08287(4) & - \\
  & cl-cl & 0.31328(12) & & 0.10664(16) & 0.7771(12) \\
  & H-cl & 0.31821(11) & & 0.09338(13) & 0.8874(12) \\
\hline
\hline
\end{tabular}
\caption{Results from the fits to $\eta_s$ meson correlators made from HISQ-HISQ, clover-clover 
and HISQ-clover $s$ quark propagators. Column 3 gives the ground-state mass in lattice units. 
The H-H and cl-cl results are very close as a consequence of tuning the bare mass parameters 
in the HISQ and clover actions. Column 4 gives the $\eta_s$ decay 
constant in lattice units for the H-H case where it is 
absolutely normalised. Column 5 gives the unnormalised 
$\eta_s$ decay constant for the cl-cl and H-cl cases. 
Column 6 gives the $Z$ factors for the cl-cl and H-cl 
cases from setting the decay constant equal to that in 
the H-H case. 
}
\label{tab:etas}
\end{table}

The decay constant of the $\eta_s$ meson can be
defined as the matrix element between the meson and the
vacuum of the temporal axial current.
When the meson is at rest this is given by 
\begin{equation}
\label{eq:fdef}
 \langle 0|A^4|\eta_s\rangle = M_{\eta_s}f_{\eta_s}. 
\end{equation}
For the HISQ action, remnant chiral symmetry gives a 
partially conserved axial current (PCAC) relation connecting the temporal 
axial and pseudoscalar currents for the Goldstone taste 
pseudoscalar that we use here. Thus we can determine an absolutely 
normalised decay constant from the relation 
\begin{equation}
\label{eq:fhisq}
f_{\eta_s} = 2m_s a_0 \sqrt{\frac{2}{E_0^3}}
\end{equation}
where $E_0$ and $a_0$ are the ground-state energy 
and amplitude respectively from the fit given 
in eq.~(\ref{eq:hisqetafit}). 
Results for the decay constant in lattice units are 
given in Table~\ref{tab:etas}. These agree with 
those from~\cite{Dowdall:2013rya} at the physical 
$s$ quark mass (see Figure 3 in that reference).  

For the clover action we do not have a PCAC relation
and so the temporal axial current must be renormalised. 
We do this by equating the decay constant obtained 
from the ground-state amplitude in the cl-cl case 
to that obtained in the H-H case where we have 
an absolute normalisation. 
In the cl-cl case we can convert the ground-state 
amplitude from our fits obtained from meson correlation functions
using the temporal axial current 
to an un-normalised decay constant value in lattice 
units using 
\begin{equation}
\label{eq:fclover}
af_{\eta_s}/Z_{A^4} =  a_{0,A^4} \sqrt{\frac{2}{E_0}}.
\end{equation}
The results of this determination are given for 
each ensemble in 
Table~\ref{tab:etas}. The renormalisation factor 
$Z_{A^4}$ is then obtained by setting $af_{\eta_s}$ 
in the cl-cl case equal to that obtained in 
the H-H case. 

An alternative method, but one that we do not use, 
would be to set the cl-cl 
decay constant equal to the physical value of 181.14(55) MeV 
obtained in~\cite{Dowdall:2013rya}. Because the discretisation 
effects seen in the H-H values of $f_{\eta_s}$ are so 
small this would make little difference  - at most 0.5\% 
on set 1. 

Exactly the same arguments and procedure apply to 
determining $af_{\eta_s}/Z_{A^4}$ and $Z_{A^4}$ in 
the H-cl case. In this case, because the $\eta_s$ 
mass is not exactly the same as the tuned value there is a 
difference between matching decay constants and 
matching matrix elements ($f_{\eta_s}M_{\eta_s}$). Because the difference in 
mass is a discretisation effect we have chosen to 
match decay constants. The differences between 
doing this and matching the matrix element $f_{\eta_s}M_{\eta_s}$ 
are as large as 6\% on set 1, but fall to 1\% on fine set 8, and act in the direction 
of making $Z_{A^4}$ smaller 
than that quoted. We can use this variation to assess the 
size of nonperturbative effects appearing in our nonperturbative determination 
of the $Z$ factors. A 6\% effect on the coarsest lattices is not a 
surprising result; $(a\Lambda)^2$ with $\Lambda$ around a few hundred MeV 
would give something similar.     

The values of $Z_{A^4}$ for the cl-cl and H-cl current are then 
given in column 5 of Table~\ref{tab:etas}. 

\begin{figure}
\centering
\includegraphics[width=0.45\textwidth]{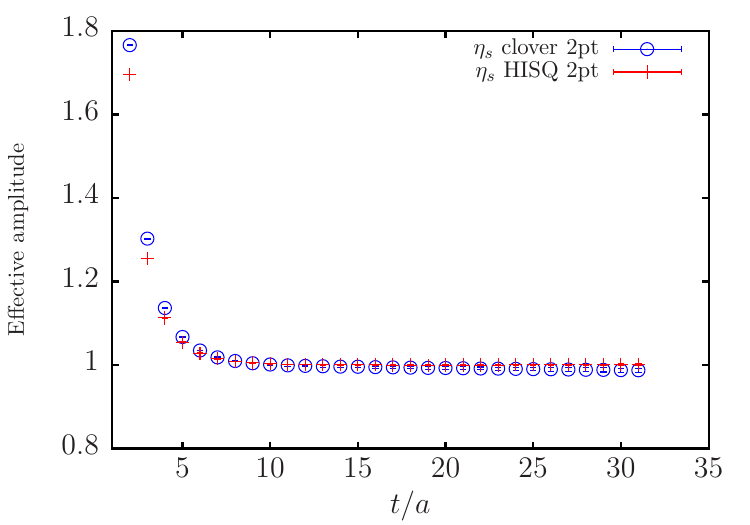}
\caption{The effective amplitude defined as the correlator divided 
by the fit result for the ground-state exponential for H-H Goldstone and 
cl-cl pseudoscalar $\eta_s$ correlators on coarse set 3. The number of 
configurations used for the H-H correlators is double 
that of the cl-cl correlators. }
\label{fig:effamp}
\end{figure}

Figure~\ref{fig:effamp} illustrates directly how similar the 
H-H and cl-cl correlators in terms of their $t$-dependence. 
The figure shows the result in each case 
of dividing 
the correlator (with the pseudoscalar current at source and sink) 
by the fit function $a_0^2f(E_0,t)$ corresponding to the 
ground state. The central value of both effective amplitudes 
is then 1 at large times. The statistical uncertainties in the H-H case 
are about 2.5 times smaller than the cl-cl case 
when double the number of configurations 
was used. 
The results for the two amplitudes agree well away 
from the central plateau region, showing that the contributions 
of excited states to the correlators are also well matched.
Discretisation errors give differences at small times. 

Another interesting feature of Figure~\ref{fig:effamp} 
is that the statistical error in the correlator 
increases with time, albeit slowly. In the simplest picture of how 
noise arises in meson correlators this would not happen
because the signal to noise ratio should be a constant 
for pseudoscalar meson correlators made of quarks with 
equal mass. The variance of the meson correlator is 
a correlator made of two quarks and two antiquarks. 
When the quark masses are the same the ground-state 
energy of this combination is twice that of the meson 
that controls the signal, in the absence of interactions 
between the two mesons and ignoring a `crossed' diagram 
that would need to be calculated to determine fully 
the two-meson correlator. It is these latter two effects 
that complicate the simple picture and cause the 
mass controlling the noise to fall below that controlling 
the signal so that an exponentially growing (albeit slowly) noise-to-signal 
ratio results. See~\cite{Gregory:2010gm, Davies:2010ip, Bazavov:2014wgs}
 for earlier discussion and 
analysis of correlator noise.  

\subsection{Z factors for $V^4$}
\label{subsec:V0}
\begin{figure}
\centering
\includegraphics[width=0.4\textwidth]{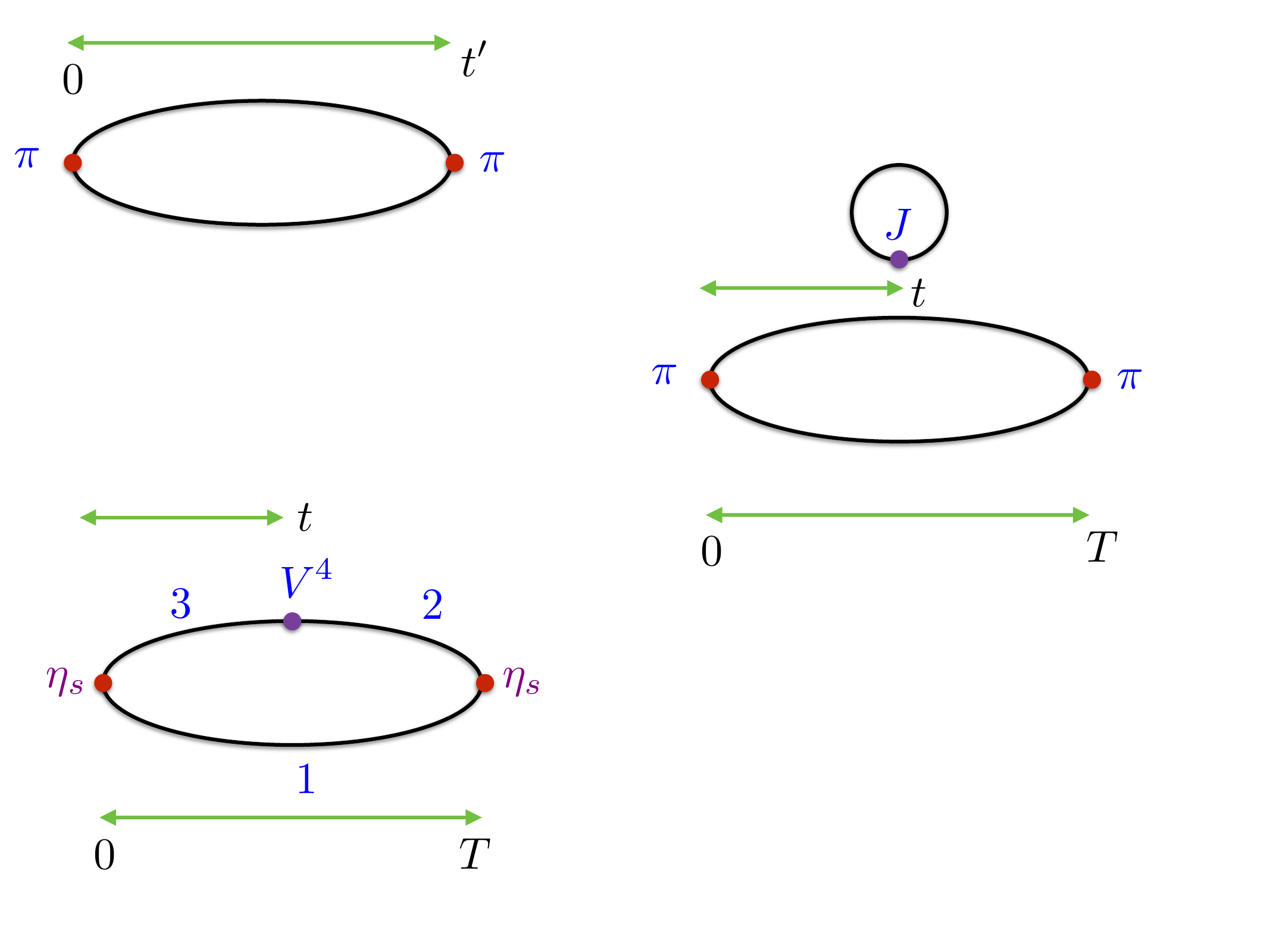}
\caption{A diagram to show how our three-point correlation functions 
are constructed. All of the quark propagators, denoted 1, 2 and 3 are for 
$s$ quarks and combined at times 0 and $T$ to make $\eta_s$ mesons. 
A temporal vector current is inserted at $t$. }
\label{fig:3pt}
\end{figure}

The normalisation of temporal vector currents in lattice QCD 
is readily obtained by demanding that the vector form factor 
be 1 between two identical states at rest. 
Here we can implement this for 
$\eta_s$ states so that
\begin{equation}
\label{eq:zvdef}
Z_{V^4} \langle \eta_s | V^4 | \eta_s \rangle =  2M_{\eta_s}.
\end{equation}
The matrix element of the vector current is calculated from 
a 3-point function as illustrated in Figure~\ref{fig:3pt}.
Propagator 2 is generated from propagator 1 as a source and 
then joined at the temporal vector vertex to propagator 3. 
Appropriate spin combinations are taken at the two ends 
to ensure that source and sink correspond to pseudoscalar mesons. 
Sums over spatial slices ensure that source and sink mesons 
are at rest. 
We use 4 values for the value of $T$ at the end-point of 
the 3-point function. This enables us to fit both 
the $t$-dependence (for $0 \le t \le T$) and the 
$T$-dependence of the 3-point function to reduce 
systematic errors from excited state contamination. 
The values used for $T$ are listed in Table~\ref{tab:valence}. 

By choosing combinations of HISQ and clover 
propagators we can determine the renormalisation factor 
for H-H, cl-cl and H-cl temporal vector currents. 
The temporal vector currents we consider are all local 
operators and for the H-H case this corresponds to 
the spin-taste structure $\gamma_4\otimes\gamma_4$.  
Because this current is not a taste-singlet we cannot 
use a three-point function made purely of staggered 
quarks but must have a non-staggered `spectator' quark 
(propagator 1 in Figure~\ref{fig:3pt}). 
Here it is natural to use a clover $s$ quark, extending our 
earlier method that used NRQCD quarks~\cite{Donald:2012ga}, itself 
based on a Fermilab Lattice/MILC method that uses clover quarks~\cite{Bazavov:2011aa}. 

\begin{figure}
\centering
\includegraphics[width=0.45\textwidth]{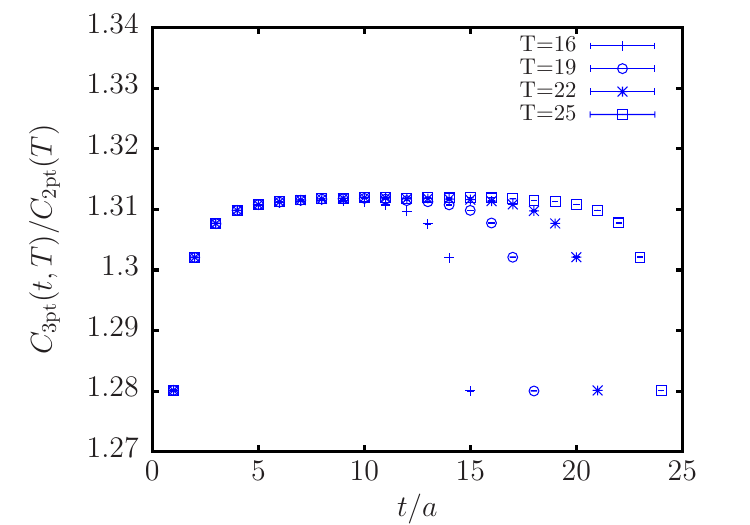}
\caption{The ratios of the average three-point correlator to average two-point correlator 
showing how the same plateau value is reached for four different values 
of T: 16,19,22 and 25 using the clover action on fine set 8. Statistical errors are 
shown on the points.}
\label{fig:3ptclo}
\end{figure}

First we discuss the case of 
 the cl-cl temporal vector current. For this case, 
all propagators are clover $s$ quarks and we use 
the pseudoscalar operator at the source and sink to make $\eta_s$ mesons. 
We make this choice because the pseudoscalar operator gives somewhat 
more precise correlators; 
the 2-point functions are simply the squared modulus of the propagator 
in that case. 
We then fit the three-point functions from all $T$ values
 simultaneously with two-point cl-cl (using $\gamma_5$ at source and sink) 
$\eta_s$ correlators. 
The fit form for the three-point function is given by:  
\begin{eqnarray}
\label{eq:simp3ptfit}
C_{3pt}=\sum_{i,j} a_iV_{ij}a_jf(E_{i},t)f(E_{j},T-t)
\end{eqnarray} 
where $a_i$ and $a_j$ are amplitudes from the two-point 
functions (eq.~(\ref{eq:hisqetafit})). We use a prior 
width on the $V_{ij}$ of 0.0(3.0) (along with priors on 
all other parameters as for the earlier two-point correlator 
fits). 
Using a relativistic 
normalisation of states the matrix element of the lattice 
temporal vector current between ground-state $\eta_s$ mesons
at rest is given by $2E_0V_{00}$ and therefore 
\begin{equation}
\label{eq:zv4}
Z_{V^4} = \frac{1}{V_{00}}.
\end{equation}
Our results for each of sets 1, 3 and 8 are listed in Table~\ref{tab:results}. 
Notice that the numbers are a lot more precise than those for $Z_{A^4}$. 
Figure~\ref{fig:3ptclo} plots the ratio of the three-point correlator 
for each value of $T$ to that of the two-point correlator at $T$, as a function 
of $t$ to illustrate 
the quality of our results. From eq.~(\ref{eq:simp3ptfit}) this ratio 
will be $1/V_{00}$ for all values of $t$, up to contamination from excited 
states. It is clear from Figure~\ref{fig:3ptclo} that this contamination is 
under good control, with all three-point functions showing a good plateau 
at the same value. Note, that we do not use this ratio in our fits, but 
instead perform a full multi-exponential fit to our correlators as 
given in eqs.~(\ref{eq:simp3ptfit}) and~(\ref{eq:hisqetafit}). 

For the H-H local temporal vector current we have a H-cl $\eta_s$ correlator 
at source and sink (made with a $\gamma_5$ operator). 
This means that there are additional oscillating terms 
in the fit form for the three-point function in a simultaneous fit with 
the appropriate two-point correlators. 
The fit function is then 
\begin{eqnarray}
\label{eq:comp3ptfit}
C_{3pt}&=&\sum_{i,j} a_iV_{ij}b_jf(E_{i},t)f(E_{j},T-t)\\
      &&  - (-1)^{(T-t)/a}\sum_{i,jo}a_iV_{ijo}b_{jo}f(E_{i},t)f(E_{jo},T-t). \nonumber \\  
      &&  - (-1)^{t/a}\sum_{io,j}a_{io}V_{ioj}b_{j}f(E_{io},t)f(E_{j},T-t). \nonumber  \\ 
      &&  + \sum_{io,jo}a_{io}V_{iojo}b_{jo}f(E_{io},t)f(E_{jo},T-t). \nonumber  
\end{eqnarray}
Again $a_i$, $b_j$, $a_{io}$ and $b_{jo}$ are amplitudes that appear 
in the two-point correlator fit (eq.~(\ref{eq:Hclsymmfit})). 
We take a prior width on $V_{ij}$ of 0.0(3.0) and on the 
other $V$ of 0.0(1.0). 
Again the renormalisation factor for the local temporal vector 
current is given by eq.~(\ref{eq:zv4}) and our values are given 
in Table~\ref{tab:results}. These results improve on the values used by 
us~\cite{Chakraborty:2014mwa, Chakraborty:2016mwy} in the calculation 
of the hadronic vacuum polarisation contribution to the anomalous 
magnetic moment of the muon. 

Lastly the H-cl temporal vector current is obtained from three-point 
functions in which propagator 3 is a HISQ quark and 1 and 2 are clover 
quarks, using the $\gamma_5$ operator to construct mesons. 
Again we fit the three-point correlators simultaneously with the 
appropriate two-point correlators. Here we need both H-cl and 
cl-cl two-point correlators. 
Our three-point function fit form has oscillatary terms on the 
side corresponding to the H-cl two-point function but none 
on the side corresponding to the cl-cl twopoint function, 
so the fit form is the first two lines of eq.~(\ref{eq:comp3ptfit}). 
We use the same priors as above 
and  again the renormalisation factor for the local temporal vector 
current is given by eq.~(\ref{eq:zv4}). Note that in using this 
equation we are ignoring small discretisation effects between 
the H-cl $\eta_s$ mass and the cl-cl $\eta_s$ mass evident 
in Table~\ref{tab:etas}. Including this effect changes the 
$Z_V$ value by less than 0.05\% even on the very coarse 
lattices. Our results are given 
in Table~\ref{tab:results}. 

\begin{table}
\begin{tabular}{llllll}
\hline
\hline
Set &  Action & $Z_{A^4}$  &  $Z_{V^4}$  &   $\rho_{{A^4}}$ & $\rho_{{V^4}}$\\ 
 &  comb'n &  &   &   & \\ 
\hline
1 & H-H  & - & 0.9881(10) & -  & -\\
  & cl-cl &0.7245(11) & 0.7262(2) & - &  -\\
  & H-cl  &0.8746(13) & 0.8660(7) & 1.0325(16) & 1.0223(9) \\
\hline 
3 &  H-H  & - & 0.9922(4) & - &  - \\
  & cl-cl  & 0.7478(9) & 0.7397(3) & - & - \\
  & H-cl   &0.8804(11) & 0.8739(7) & 1.0277(12) & 1.0201(8) \\
\hline
8 &  H-H & -  & 0.9940(5) & - & - \\
  & cl-cl &  0.7771(12) & 0.7620(3) & - & - \\
  & H-cl  & 0.8874(12)  & 0.8839(8)  & 1.0196(14) & 1.0156(10) \\
\hline
\hline
\end{tabular}
\caption{Column 4 gives results for the renormalisation factor for 
the local temporal vector current for each of the different action 
combinations and each ensemble listed in columns 1 and 2. 
Results for $Z_{V_4}$ in the H-H case are more 
precise than those given in~\cite{Chakraborty:2014mwa} becaue 
those were taken from preliminary fits.  
Column 3 repeats results from Table~\ref{tab:etas} for the 
temporal axial vector. 
Columns 5 and 6 then give the $\rho$ factors defined in Eq.~\ref{eq:zJ} 
for the off-diagonal H-cl combination for both temporal axial 
vector and temporal vector currents. Errors are statistical/fitting 
errors combined from the different components in quadrature. } 
\label{tab:results}
\end{table}

We see in Table~\ref{tab:results} that the values for 
$Z_{V^4}$ are very similar to those for $Z_{A^4}$
in the H-cl and cl-cl cases, despite being rather 
far from 1. This does add weight to the idea that 
there is a component of the $Z$ factor that comes 
from the `clover wavefunction renormalisation' and 
could be cancelled in ratios. 

\subsection{Results for $\rho_{A^4}$ and $\rho_{V^4}$}
\label{subsec:rho}

\begin{figure}
\includegraphics[width=0.45\textwidth]{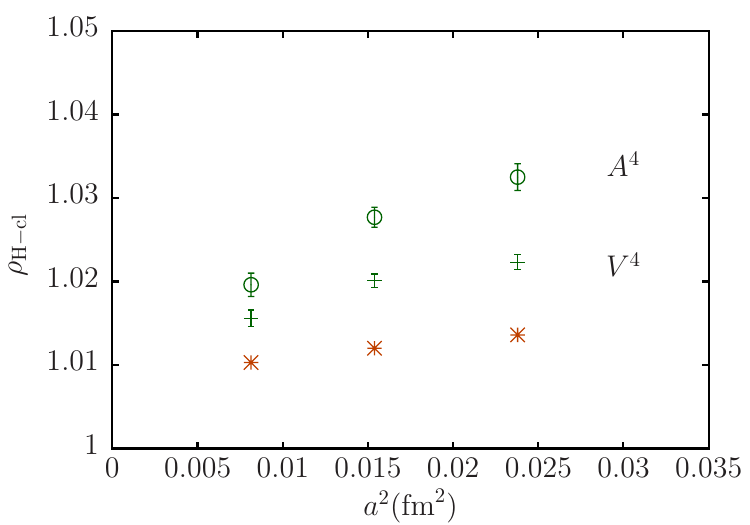}
\caption{Our nonperturbative results for the $\rho$ factors defined 
in eq.~(\ref{eq:zJ}) and given in Table~\ref{tab:results} for 
current operators made by combining HISQ and clover quarks. 
Green open circles gives results for the temporal axial 
vector current $A^4$ and green pluses the results for the temporal 
vector current $V^4$. 
Also shown are the one-loop perturbative lattice QCD 
results for mixed asqtad-clover currents 
with light clover quarks (orange bursts)}
\label{fig:rhoA4}
\end{figure}

We now have all the ingredients necessary to test the 
formula for the off-diagonal-in-action renormalisation factor 
in terms of the square root of the product of diagonal 
temporal vector renormalisation factors given in 
eq.~(\ref{eq:zJ}) (testing eq.~(\ref{eq:zhl})). 
We can do this for both temporal axial vector and temporal 
vector currents using the data in Table~\ref{tab:results}, 
and the results for the $\rho$ factors are also given 
in that table.  
The $\rho$ 
factors are indeed close to 1 in all cases, demonstrating 
that the perturbative series for $\rho$ does have small 
coefficients for all powers of $\alpha_s$. 
Note that the 
temporal vector and temporal axial vector $\rho$ factors are even 
closer to each other than they are to 1. 
In Figure~\ref{fig:rhoA4} 
we plot our values for $\rho_{A^4}$ and $\rho_{V^4}$
against the square of the lattice spacing. 

In Figure~\ref{fig:rhoA4} we
also compare to the $\mathcal{O}(\alpha_s)$ 
perturbative result for the operator made 
from a combination of clover 
and asqtad staggered quarks in the limit that both 
quark masses go to zero~\cite{Zfnalasqtad}. 
In the clover-asqtad case the $\mathcal{O}(\alpha_s)$ coefficient 
for $\rho$ for both $A^4$ and $V^4$ is $+4\pi\times 3.04\times 10^{-3} = 0.0382$ 
\footnote{Note a typographical error in~\cite{Zfnalasqtad} 
has introduced a minus sign.}, 
the same because of the chiral symmetry of the asqtad action.
In Figure~\ref{fig:rhoA4} we combine this coefficient with a value of 
$\alpha_s$ determined in the V-scheme at scale 
$2/a$ which corresponds approximately to the 
BLM scale found for these operators in the 
clover-clover case~\cite{Zfnalclover}. 
The appropriate values of $\alpha_s$ on sets 1, 3 and 8 
are: 0.356, 0.314 and 0.269. 
From these values it is clear that missing $\alpha_s^2$ terms 
in the perturbative expansion could be sizeable; a 
coefficient of 1 would give a 10\% shift to $\rho$. 

Since we are using the HISQ action for the 
staggered quark and not the 
asqtad action, the perturbative results quoted 
above are not correct for our case, and  
are provided purely for a qualitative comparison. 
However we see that the nonperturbative H-cl and 
the one-loop perturbative asqtad-cl results have similar 
values and behave in a similar way with lattice spacing. 
The nonperturbative results are slightly further 
from 1 on the coarser lattices. On the finer lattices 
they agree to within 1\%, with the perturbative result 
being 1\% from 1 and the nonperturbative result 2\%. 
Any comparison of nonperturbative and perturbative 
results must take account of possible systematic 
discretisation effects in the nonperturbative results. 
As discussed in Section~\ref{subsec:A0} we can estimate 
these from the impact of changing our definition of 
$Z_{A^4}$. This produces a sizeable 6\% effect on the 
coarsest lattices but falling to 1\% on the finest lattices.  
Thus on the finest lattices we can give an error band of 
$\pm$1\% around our 2\% difference from 1 for the $Z$ factor 
and expect the full perturbative result to lie in this band. 
If the one-loop perturbative results fall in this band, 
as the asqtad-cl results do, we can conclude that higher 
order terms in the perturbative expansion are constrained 
at this 1\% level. 

Assuming that the H-cl one-loop perturbative coeffients are similar 
to those for asqtad-cl\footnote{Preliminary indications, for which we 
thank E. G\'{a}miz, are that this is the case}, which seems likely, we 
can conclude that our nonperturbative results 
confirm the scenario in which a one-loop perturbative QCD 
determination of $\rho_J$ is a very good approximation. 
The uncertainty from missing higher 
orders in the mixed action renormalisation 
factor can then be assumed to be small on the basis 
of the known (one-loop) coefficients.  

The Fermilab-MILC asqtad-clover heavy-light calculations 
are carried out at very different values for 
the clover quark mass than that of the $s$ quark that 
we have used here. They find, however, that the 
one-loop value for $\rho$ 
varies relatively slowly with mass, becoming 
even closer to 1 as the clover mass increases 
to that of the charm quark~\cite{Zfnalasqtad}. 
Their most recent paper on $B$ meson decay 
constants~\cite{Bazavov:2011aa} with Fermilab heavy 
quarks and asqtad light quarks uses gluon field 
configurations with similar lattice spacing values to  
those used here. They take the uncertainty from missing 
higher order terms in the perturbative expansion of 
$\rho$ as 0.1$\alpha_s^2$ with
$\alpha_s$ taken as $\alpha_V(2/a)$ on the fine lattices. 
This gives a 0.7\% uncertainty from missing higher 
orders in the perturbative matching of the heavy-light 
current. 

Although at first sight this uncertainty looks very 
small for an $\mathcal{O}(\alpha_s)$ calculation we 
can see from our results that it is in fact reasonable, 
provided that the H-cl one-loop calculation gives a very similar 
result to the asqtad-cl one-loop coefficient. This 
uncertainty is compatible with the difference we see between 
our nonperturbative results and the one-loop perturbation 
theory (for asqtad-cl), allowing for discretisation effects 
in the nonperturbative results.  

In Appendix~\ref{appendix:nrqcd} we show how 
this approach to the determination of renormalisation 
constants also works when the heavy quark uses 
the NRQCD formalism. For an NRQCD-light current the 
division by the square root of the $Z$ factor 
for the vector light-light current removes sizeable 
effects in the one-loop coefficients 
associated with the light quark formalism 
for clover and asqtad light quarks; no 
such effect is present, or correction needed, 
for the NRQCD quark. 
Defining the heavy-light $Z$ factor using 
eq.~(\ref{eq:zhl}) then reduces the one-loop 
coefficient in the perturbative piece of the 
$Z$ factor from around 0.3 to around 0.05, 
with an associated reduction in perturbative 
uncertainty, given the evidence shown here.  
For NRQCD-HISQ currents the method is no longer 
useful since neither NRQCD nor HISQ has significant 
`wavefunction renormalisation' effects and the one-loop 
coefficients in $Z$ are around 0.05 without the 
use of eq.~(\ref{eq:zhl}).

We can also ask: to what extent can our results 
for $\rho$, shown in Figure~\ref{fig:rhoA4}, be 
used to constrain unknown higher order terms 
in the perturbative expansion for $\rho$?
To test this we fit a functional form to $\rho$ 
that includes a power series in $\alpha_s$ allowing 
for discretisation effects. 
We use
\begin{equation}
\label{eq:rhopert}
\rho(a,\alpha_s) = \sum_{i=0}^{n_i} \left[c_i+d_i(\frac{a\Lambda}{\pi})^2 + f_i(\frac{a\Lambda}{\pi})^4\right]\alpha_s^i 
\end{equation}
with $c_0=1.0$, $\Lambda = 0.5$ GeV and $\alpha_s$ taken in the `V' scheme at scale $2/a$. 
Priors on $c_i$, $d_i$ and $f_i$ are all taken as $0(1)$. 
Good fits (with $\chi^2/[\mathrm{dof}]$ of 0.3) 
are readily obtained to the results for both $\rho_{A^4}$ and $\rho_{V^4}$ 
with $n_i=5$ (although changing $n_i$ has very little effect). 
The fit result for $c_1$ is 0.0(1), compatible with being small, as found 
in the calculation for the asqtad-cl case~\cite{Zfnalasqtad}. The other 
$c_i$ are not constrained by the data, however. 
If instead we give $c_1$ a prior of 0.04(4) to make it close to that
for the asqtad-cl case, then $c_2$ is weakly constrained by the fit to be 
around zero with an uncertainty of 0.3. These features are again compatible 
with the perturbative series for $\rho$ having small coefficients. Given 
an $\alpha_s$ coefficient for the H-cl case, an improved constraint 
on the $\alpha_s^2$ coefficient would be possible. We show how this works in 
Appendix~\ref{subsec:Zsf} where, given an $\mathcal{O}(\alpha_s)$ coefficient, 
we are able to extrapolate the $Z_V$ results for the H-H case to finer lattices fairly 
accurately.  

\subsection{Further tests of renormalisation factors}
\label{subsec:furtherZ}

\begin{table}
\begin{tabular}{lllll}
\hline
\hline
Set & Action &  $aM_{\eta_s}$  &  $af_{\eta_s}/Z_{A^4}$ & $Z_{A^4}$ \\
& comb'n & & &  \\
\hline
1 & H-H ($\gamma_4\gamma_5\otimes\gamma_4\gamma_5$) & 0.5605(3) & 0.1409(2) &  1.0120(14) \\
\hline 
3 & H-H ($\gamma_4\gamma_5\otimes\gamma_4\gamma_5$) & 0.4396(2) & 0.1135(2) & 1.0042(18) \\
\hline
8 & H-H ($\gamma_4\gamma_5\otimes\gamma_4\gamma_5$) & 0.3157(1) & 0.08303(8) & 0.9981(11) \\
\hline
\hline
\end{tabular}
\caption{Results from the fits to $\eta_s$ meson correlators made from HISQ
$s$ quarks with the local temporal axial current operator
(in spin-taste notation $\gamma_4\gamma_5\otimes \gamma_4\gamma_5$). 
Column 3 gives the $\eta_s$ mass for this taste of meson and columns 4 and 5
the unrenormalised decay constant and derived renormalisation for this current 
as discussed in the text. 
}
\label{tab:etasnong}
\end{table}

In staggered formalisms there are multiple versions of bilinear operators corresponding 
to different `tastes'. In determing the pseudoscalar $\overline{s}s$ meson decay 
constant in Section~\ref{subsec:A0} we used the local pseudoscalar operator (in spin-taste
notation $\gamma_5 \otimes \gamma_5$) because this operator is connected to 
the partially conserved temporal axial current through the PCAC relation. 
Note that we do not actually form operators with the partially conserved temporal 
axial current because it is point-split and so quite complicated to implement. 
It is also unnecessary since we can use the simple local pseudoscalar operator. 
On some occasions, however, it is necessary or desirable to use an 
explicit temporal axial current operator. 
The simplest one is the local operator, in spin-taste 
notation $\gamma_4\gamma_5\otimes \gamma_4\gamma_5$. This couples to the `local 
nongoldstone' $\eta_s$ meson 
which has a slightly heavier mass than the goldstone meson whose mass was 
used to tune the $s$ quark mass in Section~\ref{subsec:par}.   

Here we give results for $\eta_s$ meson correlators that use this 
local temporal axial current operator at source and sink. The fits 
to these two-point correlators have staggered `oscillations' and 
we use the fit form given in eq.~(\ref{eq:Hclsymmfit}). In fact 
we fit these correlators simultaneously with the goldstone $\eta_s$ 
correlators, although the fits have no parameters in common. 
The ground-state mass, $E_0$, corresponds to the mass of the 
$\eta_s$ meson of this taste and differs from the goldstone $\eta_s$ 
mass by discretisation effects. This will be discussed further 
in Section~\ref{subsec:disc}. The ground-state amplitude, $a_0$, 
can be converted into an unrenormalised decay constant using the formula of 
eq.~(\ref{eq:fclover}). As in Section~\ref{subsec:A0} we can define a 
renormalisation constant from setting this decay constant equal 
to that obtained from the goldstone $\eta_s$ where the normalisation 
is absolute.    

Table~\ref{tab:etasnong} gives our results on sets 1, 3 and 8 for 
the mass, decay constant and $Z_{A^4}$ factor for the H-H local 
temporal axial current case. We see that $Z_{A^4}$ is very close 
to 1 on all sets. The chiral symmetry of the HISQ action 
also means that $Z_{A^4}$ for the local temporal axial vector 
current should equal that for the local temporal vector 
current~\cite{Sharpe:1993ur} up to lattice artefacts and we demonstrate that this is 
true below. 
 
Note that we would get slightly different values for 
$Z_{A^4}$ if we matched the matrix element ($f_{\eta_s}m_{\eta_s}$) 
between the tastes rather than just the decay constant. 
This is because the meson masses differ for different 
tastes by an amount proportional to $\alpha_sa^2$. Since 
this is a pure discretisation effect, we do not include it. 
Doing so would give values for $Z_{A^4}$ that are 
4\% lower on set 1 and 0.6\% lower on set 8, and in fact then 
closer to $Z_{V^4}$ on the coarser lattices.

\begin{figure}
\includegraphics[width=0.45\textwidth]{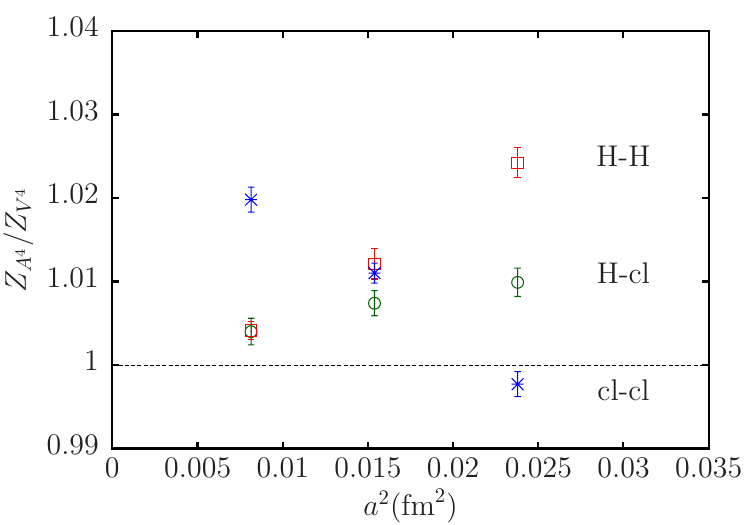}
\caption{ The ratio of renormalisation constants for local temporal 
axial and local temporal vector currents made of our 3 combinations of 
actions: H-H (red open squares), H-cl (green open circles) and cl-cl (blue 
bursts). Results are plotted as a function of the square of the lattice 
spacing and compared to 1,shown as the grey dashed line.}
\label{fig:zrat}
\end{figure}

The comparison of temporal axial vector and 
temporal vector $Z$ factors can now be done for all the combinations 
of actions we have used - H-H, H-cl and cl-cl. The HISQ action has sufficient  
chiral symmetry that the H-cl and H-H $Z$ factors should be 
the same up to lattice artefacts from the nonperturbative 
determination that vanish as $a\rightarrow 0$, and we can 
test this. For H-H 
the appropriate $Z$ factors are those for the local 
temporal vector ($\gamma_4\otimes\gamma_4$) from 
Table~\ref{tab:etasnong} and the 
local temporal axial vector ($\gamma_4\gamma_5\otimes\gamma_4\gamma_5$) 
from Table~\ref{tab:etasnong}.  For the other cases both results 
come from Table~\ref{tab:results}. 
Figure~\ref{fig:zrat} shows the ratio of $Z_{A^4}/Z_{V^4}$ as 
a function of lattice spacing. We see that, although the 
ratio differs from 1 by 2\% for H-H and 1\% for H-cl on the very coarse lattices, 
the discrepancy between the $Z$ factors for H-H and H-cl 
is falling with $a^2$ to agree to better than 1\% on 
the fine lattices. Results are consistent with the 
$Z$ factors being in complete 
agreement in the continuum limit in keeping with our expectation 
based on chiral symmetry. 

For the cl-cl case, in the absence of chiral 
symmetry, we do not expect $Z_{A^4}$ and $Z_{V^4}$ 
to agree.  In one-loop perturbation theory the 
$\mathcal{O}(\alpha_s)$
coefficient for the ratio $Z_{A^4}/Z_{V^4}$ 
is +0.127~\cite{Taniguchi:1998pf}
for the Symanzik improved gluon fields
and unimproved currents 
(along with $c_{sw}=1$ to leading order in $\alpha_s$) 
that we use here
(this is somewhat smaller than the coefficent of 0.163 for 
the unimproved gluon field case~\cite{Luscher:1996jn, Capitani:2000xi}). 
Thus we expect $Z^{A^4}/Z^{V^4}$ to be greater than 1. 
This is borne out by our results in Figure~\ref{fig:zrat}. 
Our ratio is slightly below 1 on the very coarse lattices and 
moves above 1 going towards finer lattices, heading in 
the opposite direction to the other two action combinations. 
This is consistent with results heading towards the one-loop 
perturbative result, with the discrepancy on the coarser 
lattices being mainly a result of discretisation effects.  
We have seen in Section~\ref{subsec:A0} that discretisation 
effects can be $\mathcal{O}(5\%)$ on the 
coarsest lattices used here; they would presumably be smaller had 
we used an $\mathcal{O}(a)$ improved current. 
Using the $\alpha_s$ values from Section~\ref{subsec:rho} 
would give one-loop results for $Z_{A_4}/Z_{V_4}$ of 
1.045, 1.040 and 1.034 from very coarse to fine lattices 
to be compared with the values in Figure~\ref{fig:zrat}.  
Two-loop perturbative results for $Z$ factors are available 
in the clover case~\cite{Skouroupathis:2008mf} using an 
unimproved gluon action. There including 
two-loop terms pushes $Z_{A^4}$ and $Z_{V^4}$ further below 
1 for $c_{sw}=1$  but makes less difference to their ratio.  

Ratios of renormalisation constants for two clover quarks 
are used by the Fermilab Lattice/MILC collaborations in 
their renormalisation of form factors involving a 
$b \rightarrow c$ weak transition 
(for example, $B \rightarrow D^*\ell \nu$~\cite{Bailey:2014tva}). 
In that case the two quarks 
are both heavy but of different mass and eq.~(\ref{eq:zhl}) 
is used with $l=c$. The perturbative analysis~\cite{Harada:2001fj} 
again shows very small $\mathcal{O}(\alpha_s)$ coefficients 
for the ratio $\rho$, leading to the assumption that unknown 
higher order terms are also small. In this case Heavy Quark 
Symmetry arguments can also be used in arguments about the 
size of coefficients and their mass dependence.  The results that 
we have here are for the equal mass case at small mass and so 
rather far from the $b \rightarrow c$ scenario. However the results 
for the one-loop perturbative renormalisation given above are within ~1\% of 
our nonperturbative results on the fine 
lattices (as can be seen in Figure~\ref{fig:zrat}), indicating 
that higher order corrections are indeed small in this case as in 
the H-cl case of Section~\ref{subsec:rho}. 

\subsection{Comparison of HISQ and clover discretisation effects}
\label{subsec:disc}

\begin{figure}
\centering
\includegraphics[width=0.45\textwidth]{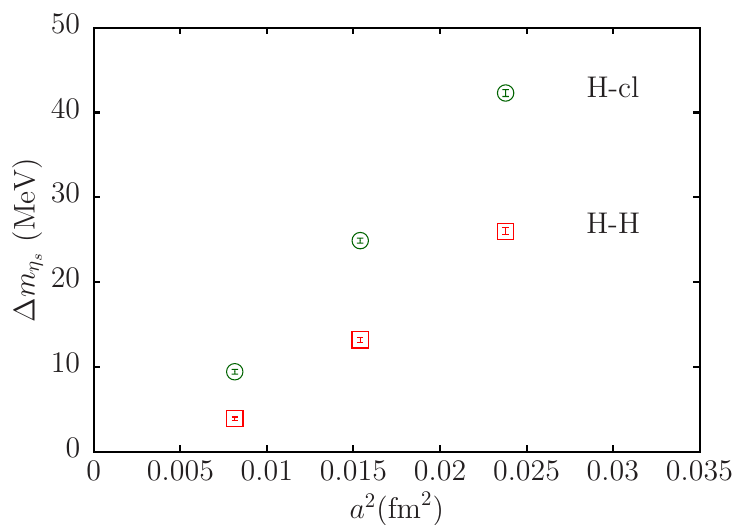}
\caption{The mass difference between the HISQ-HISQ local nongoldstone 
meson and the goldstone meson (open red squares) plotted against the 
square of the lattice spacing. Also shown is the mass difference between 
the HISQ-clover $\eta_s$ mass and that of the HISQ-HISQ goldstone 
$\eta_s$ when both HISQ and clover action are tuned to the $s$ 
quark mass (green open circles). Errors include statistical errors 
and lattice spacing uncertainties correlated between the points. 
}
\label{fig:massdiff}
\end{figure}

Systematic errors from discretisation appear differently 
in the HISQ and clover actions and we can test how much of 
an effect that is from our results. 
The first place in which discretisation effects show up is 
in differences between the masses of $\eta_s$ mesons obtained 
with two quark propagators with the quark mass tuned to that 
of the $s$ quark. 
Figure~\ref{fig:massdiff} plots two mass differences in MeV
against the square of the lattice spacing. One set of points gives 
the mass difference between the H-cl $\eta_s$ mass and that 
of the H-H goldstone $\eta_s$, using results from Table~\ref{tab:etas}. 
The second set gives the mass 
difference between two tastes of H-H $\eta_s$, the local 
nongoldstone and the goldstone, using results from Tables~\ref{tab:etas} 
and~\ref{tab:etasnong}. In both cases it is clear that 
the mass difference is purely a lattice artefact that vanishes 
as $a \rightarrow 0$. We expect the H-H mass difference to 
vanish as $\alpha_sa^2$ (since tree-level $a^2$ errors are absent from 
the action) and $a^4$. In fact for the finer two points a simple fit 
to the form $g(a\Lambda)^4+h(a\Lambda)^6$ works well with $\Lambda$ a few hundred MeV 
and $g$ and $h$ with priors $0\pm  1$; to add in the coarser 
point requires the addition of higher orders in $a^2$ and/or $\alpha_s$. 
The H-cl mass difference has
$\alpha_sa$ terms from the clover action and the results are precise enough 
to see this. A fit to the results including 
$g\alpha_s (a\Lambda)+h(a\Lambda)^2+j(\alpha_s(a\Lambda)^2)$ 
has a $\chi^2/[\mathrm{dof}]$ of 0.9. 
The H-cl mass difference is larger and has a larger slope than 
the H-H mass difference plotted in Figure~\ref{fig:massdiff}. 
It should be noted that the mass difference between the Goldstone 
and other tastes of H-H pseudoscalar meson would be larger~\cite{HISQ, Bazavov:2012uw}
than the value plotted here for the local nongoldstone to Goldstone 
splitting.    

\begin{table}
\begin{tabular}{llll}
\hline
\hline
Set & Action & $aM_{\phi}$   & $af_{\phi}/Z_V$ \\
    &  comb'n       &        &                    \\
\hline
1 & H-H & 0.8183(33) &  0.1994(33)\\
  & cl-cl & 0.7809(22) & 0.2948(33)  \\
  & H-cl & 0.8037(16) & 0.2372(16) \\
\hline
3 & H-H & 0.6475(31) & 0.1514(38) \\
  & cl-cl & 0.6306(26) & 0.2198(44) \\
  & H-cl & 0.6413(30) & 0.1789(44) \\
\hline
8 & H-H & 0.4735(13) & 0.1126(12) \\
  & cl-cl & 0.4653(14) & 0.1532(17) \\
  & H-cl & 0.4709(16) & 0.1303(13)  \\
\hline
\hline
\end{tabular}
\caption{The results for the mass and (unnormalised) decay constants 
of the $\phi$ meson in lattice units from correlators 
made of $s$ quark propagators generated using different combinations of 
HISQ and clover actions.}
\label{tab:phimf}
\end{table}

\begin{figure}
\centering
\includegraphics[width=0.45\textwidth]{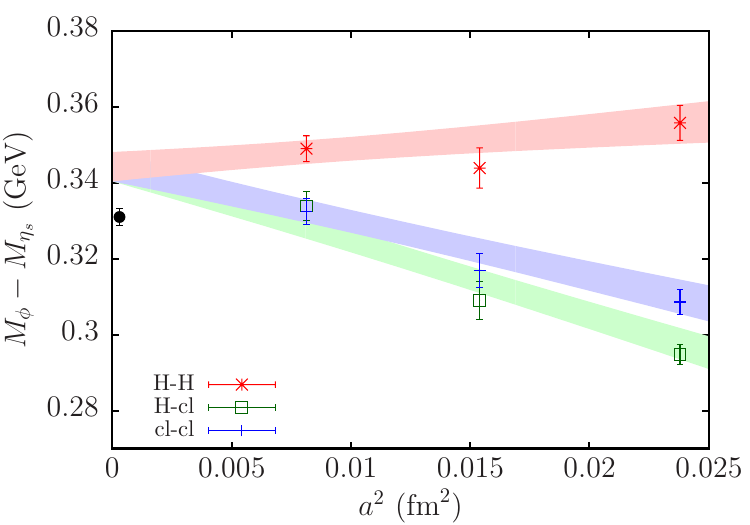}\\
\includegraphics[width=0.45\textwidth]{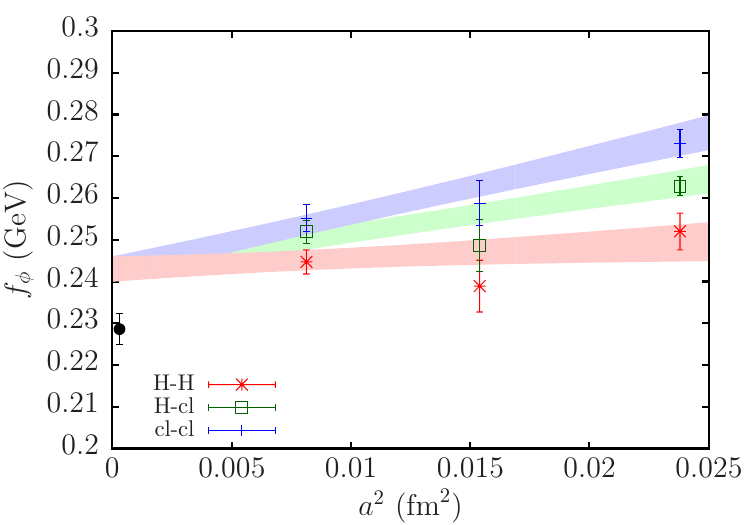}
\caption{Top: $m_{\phi}-m_{{\eta}_s}$ calculated with different 
quark formalisms and extrapolated to $a=0$. 
Red bursts give results for mesons made with two HISQ quarks, blue pluses 
those made with two clover quarks and green open squares those made 
with one HISQ and one clover quark. The associated coloured bands give 
a simple continuum extrapolation fit with a common continuum limit, 
as described in the text. The black filled circle gives the value 
corresponding to the difference of the experimental $\phi$ meson mass
the mass of the $\eta_s$ determined from lattice QCD~\cite{Dowdall:2013rya}. 
It is offset slightly from $a=0$ for visibility. 
Bottom: $f_{\phi}$ calculated for $\phi$ mesons 
made using quarks with different formalisms and 
extrapolated to $a=0$. Symbols and coloured bands are as for the top 
plot. The black filled circle is the value inferred from the 
experimental leptonic width of the $\phi$ (see text). }
\label{fig:mphifphiclh}
\end{figure}

Since we use the $\eta_s$ decay constant to 
fix $Z_{A^4}$ we cannot use that quantity to probe 
discretisation effects in the cl-cl or H-cl cases. 
That the discretisation errors are very small for 
the H-H case for this quantity has already been 
demonstrated in~\cite{Dowdall:2013rya}. 

Two further quantities that we can study to compare 
discretisation effects are the mass and decay constant 
of the vector $\overline{s}s$ state, the $\phi$. 
To reduce the impact of uncertainties in the 
lattice spacing on our results we will in fact work with 
the mass difference between the $\phi$ and the $\eta_s$.
Using the experimental value of the $\phi$ mass, 1.01946(2) GeV~\cite{pdg}, 
this difference is 0.3310(22) GeV at zero lattice spacing 
and physical quark masses, where the uncertainty comes from the 
lattice determination of the $\eta_s$ mass~\cite{Dowdall:2013rya}.

The experimental value of the $\phi$ decay constant is determined from 
its partial width to leptons using (ignoring the spread in its mass from 
its full width):  
\begin{equation}
\label{eq:decay}
\Gamma\left(\phi \rightarrow e^+e^-\right) = \frac{4\pi}{3}\alpha_{QED}^2\frac{f^2_{\phi}}{M_{\phi}}e_s^2
\end{equation}
Here $\alpha_{QED}$ at the scale of $M_{\phi}$ is $\frac{1}{137}$ and 
$e_s$ is the $s$ quark electric charge in units of e (1/3). 
The experimental value of the $\phi$ partial width 
$\Gamma(\phi \rightarrow e^+e^-) = 1.27(4) \,\mathrm{keV}$~\cite{pdg}, 
giving $f_{\phi} = 228.5\pm 3.6 \,\mathrm{MeV}$. 

We construct vector meson correlators from $s$ quark propagators 
in the same way as that described for $\eta_s$ mesons, 
combining either two HISQ propagators, two clover propagators 
or a HISQ propagator and a clover propagator. The propagators 
are combined using the spatial version of the temporal vector 
current which was normalised in Section~\ref{subsec:V0}. 
We average over all three spatial directions for the current. 
The vector meson correlators (two-point functions) are fit 
as a function of time separation between source and sink using 
the methods and fit functions outlined in Section~\ref{subsec:par}.
We use the same priors  as before;
the only difference is that now the H-H correlators have an 
oscillating component and so we use the fit form of eq.~(\ref{eq:Hclsymmfit}) rather 
than eq.~(\ref{eq:hisqetafit}). 
Table~\ref{tab:phimf} gives results in lattice units for the 
$\phi$ mass and for its un-normalised decay constant, $af_{\phi}/Z_V$,
obtained from the ground-state amplitudes returned by the fit 
according to the vector analogue of eq.~(\ref{eq:fclover})  
\begin{equation}
\label{eq:fvector}
af_{\phi}/Z_{V} =  a_{0,V} \sqrt{\frac{2}{E_0}}.
\end{equation}
To normalise the decay constant we then multiply by 
the renormalisation factor obtained for the temporal 
vector and given in Table~\ref{tab:results}, and 
by the inverse lattice spacing to convert to GeV units. 

Results are plotted as a function of the square of the 
lattice spacing for each set of action combinations 
in Figure~\ref{fig:mphifphiclh}. In order to test whether 
all the different combinations give the same continuum 
limit result, as they should, we have performed a simple 
joint extrapolation in which we allow results for each action combination 
to have a different coefficient for an $a^2$ discretisation 
effect. We also include a common term allowing for the very 
slight mistuning of the $\eta_s$ mass between lattice spacings 
and the mistuning of the sea masses from the nominal 
$m_l/m_s=0.2$ value on different ensembles. These latter effects are very small. 
Such a fit is readily achieved with a good 
$\chi^2/\mathrm{dof}$ below 0.9. The cl-cl and H-cl combinations 
in principle have $\mathcal{O}(a)$ discretisation errors 
coming from the clover quarks,
but we are not in a position to test that with our data and 
allowing for this possibility would make a joint continuum limit   
even easier to achieve. 

It is clearly visible in Figure~\ref{fig:mphifphiclh} 
that the cl-cl and H-cl combinations have larger 
discretisation effects than the H-H combination, when 
using $w_0/a$ to fix the lattice spacing. This 
is expected because the HISQ action has no tree-level $a^2$ 
errors~\cite{HISQ} so $a^2$ effects are suppressed 
by at least one power of $\alpha_s$. The clover action, even ignoring 
the possibility of $\mathcal{O}(a)$ errors, has $\mathcal{O}(a^2)$ 
errors at tree-level. We find discretisation effects for 
H-cl are about 4 times larger than for H-H in both the 
mass and decay constant. For cl-cl discretisation effects are 
3 times larger in the mass and 5 times 
larger in the decay constant, than for H-H.  

What is also seen in Figure~\ref{fig:mphifphiclh} is that the 
continuum limit of the results is not in very good agreement with 
the physical value shown as a filled black circle. 
This is because 
here we are working at unphysical $u/d$ sea quark masses. 
Better agreement will be seen in the next section where 
we map out the $\phi$ properties down to physical $u/d$ 
quark masses, but only in the H-H case. 

A further point of comparison between HISQ and 
clover quarks is that of statistical errors. These can be judged 
to some extent by looking at the fitted results for masses and 
amplitudes in the tables. We can also look directly at the 
variance of the correlators calculated on a given number of 
gluon field configurations. As already remarked in the context 
of Figure~\ref{fig:effamp} the H-H correlators that use the 
local pseudoscalar operator at source and sink have somewhat smaller 
statistical uncertainties than clover ones, even allowing for 
the different number of gluon field configurations used. 
For two-point correlation functions that use the temporal axial current, 
or that use the vector current (see Table~\ref{tab:phimf}), 
statistical errors are very similar 
between the different action combinations.  
For the determination of $Z_{V^4}$ 
using 3-point functions statistical errors are 
also similar between H-H and cl-cl on the finer sets 3 and 8 
(see Table~\ref{tab:results}; here the 
H-H and cl-cl results use the same number of gauge field 
configurations on each set). This reflects slightly lower
statistical errors on the 3-point correlators for the H-H 
current but coupled with a fit function that has also to 
account for oscillating states. 

Clover propagators are substantially more 
expensive to calculate since the Dirac matrix is an additional 
factor of 4 larger in each dimension; clover propagators are also 
16 times bigger to store. We see that the extra work associated with the 
spin degree of freedom does not lead to a reduction in statistical errors 
for the quantities that we have calculated here. 
This outcome would clearly be expected for naive quarks because 
the spin degree of freedom is then completely redundant.

\section{$\phi$ meson mass and decay constant}
\label{sec:phi}

\begin{table}
\begin{tabular}{llllll}
\hline
\hline
Set & Action & $am_s^{\mathrm{H,val}}$ & $aM_{\eta_s}$ & $aM_{\phi}$   & $af_{\phi}/Z_V$ \\
    &  comb'n & &       &        &                    \\
\hline
1 & H-H & 0.0705 & 0.54024(15) & 0.8183(33) &  0.1994(33)\\
2  & H-H & 0.0678 & 0.52652(4) & 0.7966(10) & 0.1945(8)  \\
\hline
3 & H-H & 0.0541 & 0.43134(4) & 0.6475(31) & 0.1514(38) \\
5  & H-H & 0.0533 & 0.42636(6) & 0.6385(18) & 0.1510(23) \\
7  & H-H & 0.0527 & 0.42307(2) & 0.6336(9) & 0.1507(9) \\
\hline
8 & H-H & 0.0376 & 0.31389(7) & 0.4735(13) & 0.1126(12) \\
9  & H-H & 0.0360 & 0.30484(1) & 0.4564(6) & 0.1082(6) \\
\hline
\hline
\end{tabular}
\caption{Results for the mass of the $\eta_s$ meson and 
mass and (unnormalised) decay constants 
of the $\phi$ meson in lattice units for the full set of gluon 
field configurations given in Table~\ref{tab:params} (results 
for the variable volume sets 4 and 6 will be given in Table~\ref{tab:vol}). 
Results for sets 1, 3 and 8 were already given in Tables~\ref{tab:etas} and~\ref{tab:phimf}. 
These results are all for correlators 
made of $s$ quark propagators generated using the
HISQ action only. The mass in lattice units of the valence $s$ quarks used in given 
in column 2. }
\label{tab:phiHH}
\end{table}

\begin{table}
\begin{tabular}{llllll}
\hline
\hline
Set & Action & $L_s/a$ & $am_s^{\mathrm{H,val}}$ & $aM_{\eta_s}$ & $af_{\eta_s}$  \\
    &  comb'n & &       &        &                    \\
\hline
4 & H-H & 24 & 0.0533 & 0.42664(9) & 0.11257(7) \\
5  & H-H & 32 & 0.0533 & 0.42636(6) & 0.11243(5) \\
6  & H-H & 40 & 0.0533 & 0.42642(4) & 0.11251(3)   \\
5  & H-H & 32 & 0.0507 & 0.41580(10) & 0.11122(8) \\
\hline
\hline
 &  &  &  & $aM_{\phi}$  & $af_{\phi}/Z_V$ \\
4 & H-H & 24 & 0.0533 &  0.6390(26) &  0.1504(32)\\
5  & H-H & 32 & 0.0533 &  0.6385(18) & 0.1510(23) \\
6  & H-H & 40 & 0.0533 &  0.6408(14) & 0.1526(18)  \\
5  & H-H & 32 & 0.0507 &  0.6337(17) & 0.1528(17) \\
\hline
\hline
\end{tabular}
\caption{Results for the mass and decay constant of the $\eta_s$ meson (upper table) and 
the mass and (unnormalised) decay constants 
of the $\phi$ meson (lower table) in lattice units for the sets of gluon 
field configurations of fixed $\beta$ and sea quark mass parameters 
but different spatial volume listed in Table~\ref{tab:params}. 
These results are all for correlators 
made of $s$ quark propagators generated using the
HISQ action only. The mass in lattice units of the valence $s$ quarks used is given 
in column 4. The results for $am_s^{\mathrm{H,val}}$ of 0.0533 used 1000 configurations 
from each ensemble (with 16 time sources); those for the deliberately mistuned 
value (to test tuning uncertainties) of $am_s^{\mathrm{H,val}}$ 
of 0.0507 used 300 (also with 16 time sources).  }
\label{tab:vol}
\end{table}

The fast inversion of the Dirac matrix for the HISQ 
action means that we are able to generate propagators and, consequently 
vector meson correlators, for the full set of gluon field configurations 
listed in Table~\ref{tab:params} in this case. 
By fitting the correlators, as described in Section~\ref{sec:latt}, 
we are able to determine the $\phi$ mass in lattice units and its 
decay constant using eq.~(\ref{eq:fvector}). We take results from 
6-exponential fits using a $t_{\mathrm{min}}$ value of 3 or 4, 
as for the $\eta_s$ fits. Results are given 
in Table~\ref{tab:phiHH}.   
This enables us to map out the behaviour of the $\phi$ mass 
and decay constant from values of $m_{u/d}$ in the sea of 
$m_s/5$ all the way down to their physical values and 
test the results against experiment, 
and this is what we will do here. First we discuss 
two systematic effects in the properties of the $\phi$ 
meson that we are neglecting in this calculation, and 
the impact that we expect from this in our results, 
to be included in our error budget.  

The first issue is that we have 
not included quark-line disconnected diagrams 
that would allow the $s\overline{s}$ vector to mix with 
the light isoscalar vector. Phenomenologically this is 
expected to be a very small effect, as can be seen 
from the 0.13\% branching 
fraction for the $\phi$ to decay to $\pi^0\gamma$~\cite{pdg}. 
This would be zero for a pure $s\overline{s}$ $\phi$ and 
can be 
compared to the branching fraction of 
8\% for the isospin zero light vector meson with which it can mix 
through disconnected diagrams, the $\omega$. 
There is also evidence for very small effects 
from lattice QCD calculations that have included quark-line 
disconnected diagrams.~\cite{Dudek:2013yja}  
found a mixing angle for $l\overline{l}$ in the $\phi$ 
of $1.7(2)^{\circ}$ at one value of the lattice spacing and 
a relatively heavy light quark mass. 
Analysis of quark-line disconnected correlators for 
the $s$ quark~\cite{Chakraborty:2015ugp}, albeit at much heavier sea light quark 
masses than we use here, can be used to give a 
systematic error from these missing effects and we will do that below. 

Another possible issue to worry about is the fact that the $\phi$ meson 
in the real world decays strongly to $K\overline{K}$ and 
hence is not strictly `gold-plated'. The $\phi$ meson mass 
is close to the threshold for this dominant decay, however, and 
so the $\phi$ width is rather small at 4 MeV~\cite{pdg}. 
A simple model suggests 
that coupling to the $K\overline{K}$ might contribute -5 MeV to 
the $\phi$ mass~\cite{Donald:2013pea} in the continuum. 
We expect lattice QCD calculations to be able to reproduce the 
$\phi$ meson mass to this level of accuracy then, even if 
the coupling to the $K\overline{K}$ decay mode is distorted 
on the lattice. 

In lattice QCD calculations the $\phi$ is stable for two reasons. 
The first is that the $K$ meson mass depends on the $u/d$ 
quark mass and so is heavier than its physical value when 
the $u/d$ are unphysically heavy. We can explore this issue 
here because we have results for a wide range of $u/d$ quark masses.
Note that, in the absence of coupling to $K\overline{K}$, we would 
expect very little $u/d$ quark mass-dependence for the properties 
of the $\phi$, comparable with that seen for the 
$\eta_s$ decay constant (for fixed $\eta_s$ mass) 
and mapped out in~\cite{Dowdall:2013rya}. 
The second reason for $\phi$ stability 
is that the $\phi \rightarrow K\overline{K}$ decay proceeds 
via a P-wave because the $\phi$ has spin 1; 
a zero momentum $\phi$ 
must decay to 2 $K$ mesons of equal and opposite 
non-zero momentum. 
In the continuum the non-zero momentum can be arbitrarily small, 
but the minimum lattice spatial momentum is $2\pi/L_s$. 
The experimental $\phi$ and 
$K$ meson masses would require a lattice spatial extent of
$L_s \approx$ 10 fm for the energy of the decay products 
to fall below the $\phi$ mass. This is almost double the size 
of the largest lattice that we use, typical of state-of-the-art 
lattice QCD calculations. So in practice this means that 
$\phi$ mesons are always stable on the lattice.  

We have tested the dependence of the $\phi$ meson 
mass and decay constant on the lattice volume for one set of 
simulation parameters, that corresponding to gluon configuration sets 
4, 5 and 6 given in Table~\ref{tab:params}. These sets have the  
same lattice spacing, $a\approx 0.12$ fm, and 
$m^{\mathrm{sea}}_{u/d}= m^{\mathrm{sea}}_s/10$. 
Their lattice volumes differ from 24 points on a side ($L_s \approx 3$ fm) 
to 40 points on a side ($L_s \approx 5$ fm). The $\phi$ mass 
and decay constant, and those of the $\eta_s$, are given 
in Table~\ref{tab:vol}.  We see that, within the $0.2 - 0.4$\% 
statistical uncertainties that we have, there is no significant 
effect of the lattice size on the $\phi$ mass.  This is also 
true for the decay constant within the larger $1 - 2$\% uncertainties 
that we have in that case. A further test comes from the fact that we can 
fit the independent results on the 3 ensembles simultaneously 
demanding that they give the same fit parameters for energies 
and amplitudes and obtain a good fit. 

Although we have not calculated the $K$ mass 
here, we can estimate its value accurately from results at similar 
masses in~\cite{Dowdall:2013rya}. This gives a (Goldstone) $K$ mass in lattice 
units of 0.315 for the valence $s$ quark mass used here and a $u/d$ quark 
mass given by that in the sea, 
so that $2M_K < M_{\phi}$. 
However, the value of twice the energy of a $K$ meson with the 
minimum lattice momentum would vary, in lattice units, from 0.820 on the $24^3$ 
lattice (set 4) to 0.704 on the $40^3$ lattice (set 6). 
The values of $2E_K^{\mathrm{min}}$ on all of the volumes 
are then more than 100 MeV above the corresponding mass 
of the $\phi$. 
In fact, for staggered quarks, $2E_K^{\mathrm{min}}$ would be somewhat higher 
than these estimates 
because the $\phi$ that we use here cannot decay to 2 Goldstone-taste 
$K$ mesons. Instead we must sum over different appropriate pairs 
of tastes~\cite{Chakraborty:2016mwy}, all of which have masses that 
are heavier 
than the Goldstone by an $\mathcal{O}(a^2)$ effect. 
This then increases further, typically by 50 MeV on these coarse lattices, 
the discrepancy between $M_{\phi}$ and 
$2E_K$. The finite-volume 
impact of coupling between $\phi$ and $K\overline{K}$ is then not visible
with our statistical accuracy, because $2E_K$ is too far 
above $M_{\phi}$. 

The only significant finite volume effect that we see in 
Table~\ref{tab:vol} is that in the mass of the $\eta_s$ on 
the smallest, $24^3$, lattices (set 4).  At 0.06\% the effect 
is tiny but somewhat larger than the $\mathcal{O}(0.01\%)$ that 
might have been expected from NLO chiral perturbation 
theory~\cite{Dowdall:2013rya}. 
A similar effect is seen in $aM_{\pi}$ in~\cite{Bazavov:2013vwa}.  
Note however that no significant difference is seen between 
results on the $32^3$ and $40^3$ lattices. These lattices
have sizes in 
units of $M_{\pi}$ of $M_{\pi}L_s > 4$, more typical of the 
other ensembles used here. 

\begin{table}
\begin{tabular}{lll}
\hline
\hline
Error & $M_{\phi}$ &  $f_{\phi}$ \\
\hline
statistics & 2.9 & 2.0\\
$Z_V$ & - & 1.2 \\
$a^2 \rightarrow 0$  & 2.1 &  1.7 \\
$m_{u/d}$ tuning  & 0.1 &  0.1 \\
$m_{s}$ tuning  & 0.4 &  0.4 \\
$M_{\eta_s}$ value & 2.2 & - \\
$K\overline{K}$ decay & 2.5 & 0.7  \\
`disconnected' diagrams & 4.0 & 0.6 \\
\hline
Total & 6.3 & 3.1\\
\hline
\hline
\end{tabular}
\caption{Error budget for our results for the mass 
and decay constant of the $\phi$ meson. 
Contributions to the error are given in MeV.
The uncertainty from the value of $M_{\eta_s}$ feeds into 
$M_{\phi}$ (only) since the fitted quantity used to derive 
$M_{\phi}$ is the mass difference. The $m_s$ tuning uncertainty 
comes from the fit, using a deliberately mistuned $s$ quark 
mass to assess the impact of the accuracy of our tuning 
on the quantity being fitted. 
 }
\label{tab:err}
\end{table}

\begin{figure}
\centering
\includegraphics[width=0.45\textwidth]{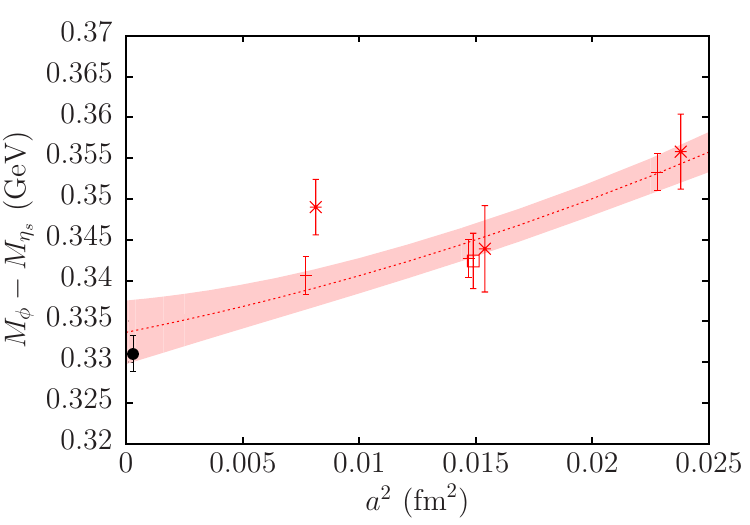}\\
\caption{Results from Table~\ref{tab:phiHH} for 
$M_{\phi}-M_{{\eta}_s}$ calculated with 
the HISQ action on a wide range of gluon field configurations 
and plotted against the square of the lattice spacing. 
Red bursts give results for $u/d$ quark masses equal 
to $m_s/5$ (sets 1, 3 and 8), red open squares for 
$m_{u/d} = m_s/10$ (set 5), and red pluses for $m_{u/d}$ 
close to its physical value (sets 2, 7 and 9). 
Note that the result for a mistuned $s$ mass, included 
in the fit, is not plotted. 
Error bars include statistical errors and uncertainties 
from the lattice spacing that are correlated between 
the points. 
The red shaded band and dotted red line give the result of a chiral/continuum 
fit described in the text, evaluated at physical 
$m_{u/d}$  as a function of lattice spacing. 
The black filled circle gives the value 
corresponding to the difference of the experimental $\phi$ meson mass
the mass of the $\eta_s$ determined from lattice QCD~\cite{Dowdall:2013rya}. 
It is offset slightly from $a=0$ for visibility. }
\label{fig:mphietahisq}
\end{figure}
 
We now move on to look at how the $\phi$ meson mass is 
affected by the $u/d$ quark mass in the sea. Our results, 
in Table~\ref{tab:phiHH}, include 
$u/d$ quark masses from $m_s/5$ down to the physical 
value ($m_s/27.4$~\cite{pdg}). The spatial size of the 
lattices, $L_s$, is approximately constant in units of 
$M_{\pi}$ with $M_{\pi}L_s$ values varying from 3.3 to 
4.6~\cite{Dowdall:2013rya}. The minimum energy of 
virtual $K\overline{K}$ pairs then falls linearly with 
$m_{u/d}$ towards the physical point 
both as $M_K$ falls and as the minimum spatial momentum 
falls. We might then expect to see some impact on $M_{\phi}$ 
from changing $m_{u/d}$. As an example, on the fine physical 
point lattices (set 9)
the minimum $2E_K$ for the Goldstone $K$ meson 
is only 40 MeV above $M_{\phi}$. Staggered taste-effects, 
reduced by over a factor of two compared to the coarse lattices 
discussed above, 
typically gives a further 20 MeV. The impact of taste-effects 
means that we need to allow for $a^2$-dependent $m_{u/d}$ 
effects in our fits used to determine the physical 
(continuum and chiral) limit of our results, and we will do this below.

Figure~\ref{fig:mphietahisq} shows our results for the difference of the 
$\phi$ and $\eta_s$ masses as a function of lattice 
spacing. We use the difference, as we did in Section~\ref{subsec:disc},
rather than the $\phi$ mass itself, to reduce uncertainties 
from the lattice spacing\footnote{A fit to the ratio $M_{\phi}/M_{\eta_s}$ 
also avoids large lattice spacing uncertainties but the statistical errors 
in the $\phi$ mass lead to a larger uncertainty in $M_{\phi}$ at the 
physical point.}.
The different symbols indicate 
results at different values of the $u/d$ quark mass. We see that 
on the fine lattices there seems to be a difference between   
results at $m_{u/d}/m_s = 1/5$ (red burst) and $m_{u/d}$ at its physical 
value (red plus), whereas there is no clear difference on 
the very coarse lattices. This is consistent with the expectation 
above, but is not very significant given our statistical uncertainties. 

To extract a physical result we fit the results to a simple 
functional form in $a^2$ and $m_{u/d}$, allowing for correlations 
between the points coming from the determination of the lattice 
spacing. The functional form that we use is: 
\begin{eqnarray}
[M_{\phi}&-&M_{\eta_s}](a,m_{u/d}) = [M_{\phi}-M_{\eta_s}]_{\mathrm{phys}}\times \nonumber \\
&& \left[ 1 + c_{a^2}(\Lambda a)^2  + c_{a^4}(\Lambda a)^4 + c_{a^6}(\Lambda a)^6 \right. \nonumber \\
&& + c_{\delta}\frac{\delta m}{10}(1+c_{\delta a^2} (\Lambda a)^2) \nonumber \\
&& \left. + c_{s}(M_{\eta_s}-0.6885 \,\mathrm{GeV})\right] . 
\label{eq:mfit}
\end{eqnarray} 
Here $[M_{\phi}-M_{\eta_s}]_{\mathrm{phys}}$ is the physical value in the continuum and chiral limit; 
we take a prior of 0.3(1) on this value. Coefficients $c_{a^n}$ allow for 
discretisation effects; we take priors of 0.0(1.0) on these values, except 
for $c_{a^2}$ for which we take 0.0(0.5) since there are no tree-level $a^2$ errors 
in the HISQ action~\cite{HISQ}. In fact the higher order terms, $c_{a^4}$ and 
$c_{a^6}$, have little impact on the fit. $c_{\delta}$ allows for the effect of unphysical 
$u/d$ quark masses and $c_{\delta a^2}$ for $a^2$-dependence in these effects. 
Here $\delta m$ is difference of $2m^{\mathrm{sea}}_{u/d}+m^{\mathrm{sea}}_s$ and 
its tuned value in units of the tuned $s$ quark mass~\cite{Chakraborty:2014aca}. 
Dividing by 10 converts it into a chiral scale.  We take very wide priors of 0.0(5.0) 
on $c_{\delta}$ and $c_{\delta a^2}$ to allow for the effects of $K\overline{K}$ 
coupling to the $\phi$ giving more pronounced dependence than is normally 
seen in gold-plated meson masses. In fact the width of this prior makes little 
difference to the physical point result. Finally, $c_s$ allows for slight mistunings 
of the $s$ quark mass, as measured by mistuning of $M_{\eta_s}$. 
Here we make use of the results given in Table~\ref{tab:vol} at a deliberately 
mistuned valence $s$ mass of 0.0507 to estimate this parameter and include these 
results to enable it to be fixed within 
the fit. We take the prior on $c_s$ of -0.5(0.5).  

The fit gives a $\chi^2/\mathrm{dof}$ of 0.97 for 8 degrees of freedom 
(the 7 tuned $s$ mass data points plus the mistuned value). 
The fitted curve evaluated at the physical sea quark masses ($\delta m = 0.0$) 
is plotted as a red band in Figure~\ref{fig:mphietahisq}. 
The physical result is 0.335(4) GeV in good agreement with the value expected 
from the experimental $\phi$ mass of 0.331(2) GeV. This is a significant 
improvement on our earlier value~\cite{Donald:2013pea} using gluon field
configurations that include 2+1 flavours of asqtad quarks in the sea 
but at heavier-than-physical $u/d$ quark masses. 
Adding back in the $\eta_s$ meson mass, with its 2.2 MeV uncertainty, gives 
a lattice QCD result of 
\begin{equation}
\label{eq:mphi}
M_{\phi} = 1.0232(42)(25)(40)\, \mathrm{GeV} 
\end{equation} 
to compare to 
the experimental result of 1.0195 GeV~\cite{pdg} 
(with a sub-MeV uncertainty). Here the second error of 2.5 MeV 
is included to allow for the incomplete treatment of the $K\overline{K}$ 
decay mode. We take this as half the expected shift in $M_{\phi}$ from 
coupling to $K\overline{K}$, given that there is 
evidence in our results of sea $u/d$ quark 
mass-dependence consistent with some impact from this effect. 
The third error, of 4 MeV, allows for the missing quark-line 
disconnected correlators. In~\cite{Chakraborty:2015ugp}, the same result 
for the $\phi$ mass was found, to an accuracy of 0.4\%, whether 
quark-line disconnected diagrams were included in the fit or not, so 
we take this as the uncertainty. 
Our error budget is given in Table~\ref{tab:err}. 

\begin{figure}
\centering
\includegraphics[width=0.45\textwidth]{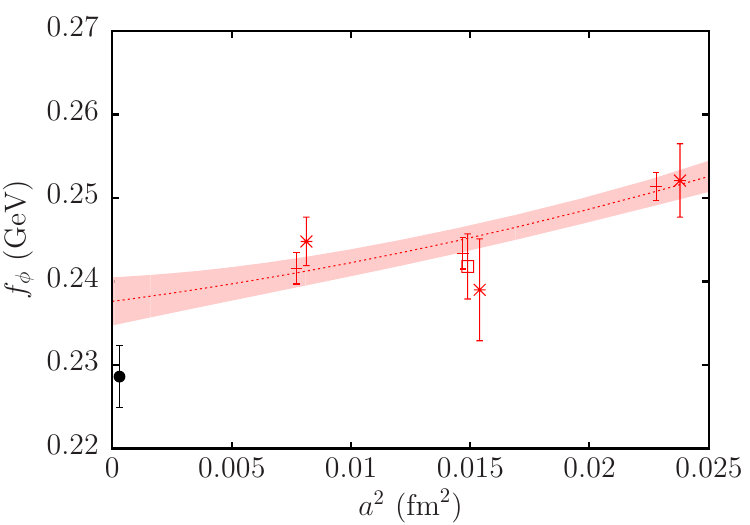}\\
\caption{Results from Table~\ref{tab:phiHH} for 
$f_{\phi}$, the $\phi$ meson decay 
constant, calculated with 
the HISQ action on a wide range of gluon field configurations 
and plotted against the square of the lattice spacing. 
Red bursts give results for $u/d$ quark masses equal 
to $m_s/5$ (sets 1, 3 and 8), red open squares for 
$m_{u/d} = m_s/10$ (set 5), and red pluses for $m_{u/d}$ 
close to its physical value (sets 2, 7 and 9). 
Note that the result for a mistuned $s$ mass, included 
in the fit, is not plotted. 
Error bars include statistical errors and uncertainties 
from the lattice spacing and current renormalisation 
factor $Z_V$ that are correlated between 
the points. 
The red shaded band and dotted red line give the result of a chiral/continuum 
fit described in the text, evaluated at physical 
$m_{u/d}$  as a function of lattice spacing. 
The black filled circle gives the value 
inferred from the experimental width for $\phi$ decay to $e^+e^-$~\cite{pdg}. 
It is offset slightly from $a=0$ for visibility. }
\label{fig:fphihisq}
\end{figure}

Our analysis of the $\phi$ meson decay constant proceeds in a similar
way to that of the mass. Figure~\ref{fig:fphihisq} plots the 
results from Table~\ref{tab:phiHH} as a function of lattice spacing. 
To convert the results for $af_{\phi}/Z_V$ in Table~\ref{fig:fphihisq}, obtained directly from 
the ground-state amplitudes of fits to our 2-point correlators, into results 
in physical units for $f_{\phi}$ we need to multiply by $a^{-1}$ in GeV from
Table~\ref{tab:params} and the current renormalisation, $Z_V$ from Table~\ref{tab:results}. 
We have determined $Z_V$ on only one ensemble from each group with almost the 
same lattice spacing. However, we do not expect $Z_V$ to vary significantly 
between for example, sets 3, 5 and 7. Physically $Z_V$ relates currents 
between two different regularisations of QCD (the continuum and the lattice) 
that differ in their ultraviolet modes. It can be expressed in QCD perturbation 
theory (although we have chosen to calculate it nonperturbatively)
as a power series in $\alpha_s$ where the scale of $\alpha_s$ is related 
to the inverse lattice spacing. The $Z_V$ values we have for the HISQ 
case are very close to 1, with a difference of 1 of about 0.01. 
Assuming this comes purely from an $\mathcal{O}(\alpha_s)$ term we 
can estimate the effect on $Z_V$ of the small changes in $a$ between 
sets 3 and 7 of 2\%. This gives an expected change in $Z_V$ of 0.0001, 
smaller than our uncertainties. 
We therefore use the $Z_V$ results from Table~\ref{tab:results}
to renormalise the results from all of our ensembles, including the 
uncertainty from $Z_V$, and its correlation between sets of results, 
in our continuum/chiral fit. 

Our continuum and chiral fit for $f_{\phi}$ takes exactly the 
same form as that given in Eq.~\ref{eq:mfit} and has the same priors, 
except for the prior on the physical result which we now take to 
be 0.2(0.1) and on $c_s$ which we take as 0.0(0.5) as there is no 
phenomenological reason to expect a strong dependence in either direction for 
the effect of mistuning the $s$ quark mass.  
Our fitted curve evaluated at physical sea quark masses is shown 
as the shaded band on Figure~\ref{fig:fphihisq}. The physical 
result that we obtain (with a $\chi^2/{\mathrm{dof}}$ of 0.71 
for 8 degrees of freedom) is 
\begin{equation}
\label{eq:fphires}
f_{\phi} = 0.2376(29)(7)(6) \,\mathrm{GeV}.
\end{equation}
This is in reasonable agreement (within 2$\sigma$) of the result 
of 0.2285(36) GeV inferred from experiment (see eq.~(\ref{eq:decay})) 
and again is a signficant improvement on our earlier 
result~\cite{Donald:2013pea}.  
The second uncertainty here is an estimate of the 
impact of coupling to the $K\overline{K}$ decay mode. 
We take an estimate of 20 MeV for the difference of 
$\rho$ and $\omega$ decay constants inferred from their 
leptonic decay rates and multiply by the ratio of 
$\phi$ to $\rho$ total widths~\cite{pdg} (4/150) and the ratio of 
$\phi$ to $\omega$ decay constants (0.23/0.2). The $\rho$ decays strongly to 
$\pi\pi$ but the $\omega$, having isospin zero, cannot do 
this. The large width of the $\rho$ makes the determination 
of its decay constant problematic but a 20 MeV difference 
between $\rho$ and $\omega$ results from the application 
of eq.~(\ref{eq:decay})~\cite{Chakraborty:2015ugp}, remembering 
to allow for the isospin difference between the two mesons that 
reduces the `effective charge' in the $\omega$ case to one third 
that of the $\rho$. 
If we assume that the 20 MeV is an 
indication of the size of effects from strong decays then 
reducing this in proportion to the total width of 
$\rho$ and $\phi$ and increasing in proportion to the decay 
constant
gives an estimate of 0.6 MeV 
for the impact on the $\phi$. 
The third uncertainty, of 0.7 MeV, is an estimate of the 
impact of missing quark-line disconnected diagrams. We obtain 
this from the analysis in~\cite{Chakraborty:2015ugp}, where the 
effect of the disconnected $s\overline{s}$ correlator on the anomalous 
magnetic moment of the muon was found 
to be -0.05\% of that of the light-quark connected correlator. This is 
equivalent to -0.5\% of the $s$-quark connected correlator (for the light 
quark masses used there). 
Taking the connected correlator contribution to be approximately 
proportional to the square of the decay constant~\cite{Chakraborty:2015ugp} 
implies a -0.3\% effect on the decay constant. Rather than take 
this as a one-sided error, we simply take the value as an estimate 
of the uncertainty.

\section{Conclusions}
\label{sec:conclusions}

The results presented here range over an apparently rather broad set 
of topics but they are all linked through the necessity to reduce 
and to test uncertainties obtained from lattice QCD calculations 
for decay rates that can be compared to experiment. 
We have focussed here on mesons made from valence $s$ quarks because, 
although the $s$ quark is light in QCD terms $s$ quark propagators 
are considerably faster to generate in lattice QCD than those 
containing $u/d$ quarks and correlators are statistically more precise, 
enabling systematic effects to be more clearly seen.  
We work on state-of-the-art gluon field configurations that include 
the effect of $u$, $d$, $s$ and $c$ quarks in the sea with an 
improved gluon action to minimise systematic discretisation effects 
coming from anywhere other than the different quark actions that 
we compare. 

Our first analysis here has compared renormalisation constants, 
determined nonperturbatively, for temporal axial vector and 
temporal vector currents constructed either from HISQ quarks
or from clover quarks or, in a mixed-action approach, from one 
clover and one HISQ quark. For the temporal axial current 
case we have used the fact that pseudoscalar correlators 
made from HISQ quarks can be absolutely normalised. 
For the temporal vector case we have used the fact that 
the vector form factor between two identical mesons at 
rest should be 1. Our results show that the renormalisation
constants for the clover-clover case are very different 
from 1, as expected from one-loop perturbation theory, but 
that the mixed-action currents inherit elements of this 
renormalisation in a relatively simple way, as suggested by 
the work of~\cite{Zfnalclover}. This means that the ratio of 
the mixed action renormalisation constant to the square 
root of the product of the renormalisation constants for 
the local temporal vector currents for the unmixed action 
cases (i.e. $\rho$ in eq.~(\ref{eq:zJ})) 
is close to 1. Our nonperturbative test of this 
relationship means that it is indeed valid to calculate 
this ratio to one-loop in lattice QCD perturbation theory 
and take a small uncertainty (of $\mathcal{O}(1\%)$) 
from missing higher order terms, as the Fermilab Lattice/MILC 
collaborations do in their work on $B$ and $D$ meson 
decay constants~\cite{Bazavov:2011aa} and $B\rightarrow \pi \ell \nu$ 
form factors~\cite{Lattice:2015tia} using a mixed clover-staggered 
approach. Thus our results provide 
confirmation, after the fact, of this element of their 
error budget. In Appendix~\ref{appendix:nrqcd} we show a 
similar perturbative analysis for mixed NRQCD-light currents, 
justifying the normalisation element of the error budget 
in the $B$ decay constant~\cite{Dowdall:2013tga} and 
$B \rightarrow \pi \ell \nu$ 
calculations~\cite{Colquhoun:2015mfa} in this case. 

Modifications to the Fermilab heavy quark approach have been 
used in the Relativistic Heavy Quark (RHQ) formalism~\cite{Christ:2006us} 
by the RBC/UKQCD collaboration~\cite{Christ:2014uea, Flynn:2015mha}. 
The modifications involve tuning 
some coefficients nonperturbatively to reduce leading systematic 
errors. The approach to the normalisation of heavy-light currents 
is the same, however, using eq.~(\ref{eq:zhl}) to define 
the ratio $\rho$ and then determining $\rho$ to one-loop in 
lattice QCD perturbation theory. The coefficient of $\alpha_s$ in 
$\rho$ is somewhat larger for the RHQ-domain wall current than in the 
Fermilab-asqtad case, but it is still numerically small at ~0.1~\cite{Christ:2014uea} 
for the temporal axial current. 
The uncertainty in $f_B$ and $B \rightarrow \pi$ form factors from missing 
higher-order terms in the perturbative expansion is 
taken as the size of the one-loop term in $\rho$, arguing that, 
as for the Fermilab case, the one-loop term is indicative of what 
will appear at higher orders.  
Our results are not directly applicable to this case and it 
is harder to argue about the `natural' size of perturbative coefficients 
when there are relatively large nonperturbative coefficients 
(such as that of the clover term) in the action. It would 
be straightforward to provide a consistency check by 
repeating the analysis that we have done here, substituting 
a light RHQ field for the clover quark and a domain-wall 
quark for the HISQ quark. 

Such tests are important because lattice QCD 
determination of these decay constants and form factors 
feeds into determination of CKM elements such as $V_{ub}$ 
through comparison with experimental exclusive decay modes. 
Accuracy on CKM elements is critical to over-constraining 
the Standard Model in the search for new physics. Currently 
the discrepancy in $V_{ub}$ determination using inclusive 
and exclusive processes is a cause for concern~\cite{pdg} 
and resolution will require improved accuracy from both 
determinations. On the exclusive side, we need to be sure 
that we understand sources of uncertainty in the lattice 
QCD calculation and our result here provides reassurance that 
we do understand uncertainties from current normalisation. 

Our further analysis has focussed on the mass and 
decay constant of $\phi$ mesons, using the vector current 
renormalisation factors to fix the normalisation of 
the decay constant. We have seen that the results from 
all three possibilities, using the HISQ action or the clover 
action or the mixed-action approach, agree in the continuum 
limit as they should on a set of ensembles with a fixed 
heavier-than-physical $u/d$ quark mass. This is an important 
and independent consistency check of our results and they show, 
as expected, larger discretisation effects with the clover action  
than with the more highly improved HISQ action. 

To study the mass and decay constant of the $\phi$ meson 
closer to the physical point, we have used the HISQ action 
(only) on a wider set of gluon field configurations that include 
different values of the $u/d$ quark masses in the sea going 
down to the physical value. 
We include single-meson quark-line connected diagrams only 
since we believe, based on phenomenological evidence, that 
the impact of quark-line disconnected diagrams and 
coupling of the $\phi$ to its $K\overline{K}$ decay mode, which 
is virtual on the lattice, 
is small. We may be seeing some evidence of the effect of this coupling 
in enhanced dependence of the $\phi$ meson mass on 
the $u/d$ sea quark mass. Our final results are:  
\begin{eqnarray}
\label{eq:finalphi}
M_{\phi} &=& 1.0232(42)(25)(40)\, \mathrm{GeV} \nonumber \\ 
f_{\phi} &=& 0.2376(29)(7)(6) \,\mathrm{GeV}.
\end{eqnarray}
The second error in both cases is an estimate
of the remaining effect of the $K\overline{K}$ mode, 
and the third error, an estimate of the impact of 
missing quark-line disconnected diagrams. 
Our results are in good agreement with experiment 
and the $\mathcal{O}(5\,\mathrm{MeV})$ uncertainties 
are a significant improvement on earlier results. 
The accuracy of our $\phi$ meson correlators 
led to the first flavour-separated determination 
of the valence $s$ quark hadronic vacuum polarisation 
contribution to the anomalous magnetic moment 
of the muon~\cite{Chakraborty:2014mwa}; an uncertainty 
of 1\% was reached in that calculation. These 
uncertainties 
are also promising for improvements to lattice QCD calculations 
of form factors for decay processes that 
include $\phi$ mesons~\cite{Horgan:2013pva,Donald:2013pea}. 

\subsection*{\bf{Acknowledgements}} 

We are grateful to the MILC collaboration for the use of their configurations and to 
R. Dowdall, A. El-Khadra, E. G\'{a}miz, A. Kronfeld and R. van de Water 
for useful discussions. 
Computing was done on the Darwin supercomputer at the University of 
Cambridge as part of STFC's DiRAC facility. 
We are grateful to the Darwin support staff for assistance. 
Funding for this work came from the Gilmour bequest to the University of 
Glasgow, the National Science Foundation, the Royal Society, 
the Science and Technology Facilities Council 
and the Wolfson Foundation. B. C. is supported by the U.S. Department 
of Energy Office of Science, Office of Nuclear Physics under 
contract DE-AC05-06OR23177. 

\appendix

\section{Renormalisation of NRQCD-light currents}
\label{appendix:nrqcd}

\begin{table}  
\begin{tabular}{llll}
\hline
\multicolumn{4}{c} {NRQCD-clover} \\
\hline
$Ma$ & $n$ & $z_0^{(1)}$ & $\rho^{(1)}$ \\
\hline
4.0 & 2 & -0.2972 &  -0.0077     \\ 
3.0 & 2 & -0.3533 &  -0.0638     \\ 
2.0 & 2 & -0.3002 &  -0.0107    \\
1.2 & 3 & -0.2096 &  +0.0799     \\
\hline
\multicolumn{4}{c} {NRQCD-asqtad} \\
\hline
$Ma$ & $n$ & $z_0^{(1)}$ & $\rho^{(1)}$ \\
\hline
4.0 & 2 & 0.272 & 0.067      \\ 
2.8 & 2 & 0.209 &  0.0035     \\ 
1.95 & 4 & 0.154 &  -0.052     \\
1.2 & 6 & 0.154 &  -0.052     \\
\hline
\multicolumn{4}{c} {NRQCD-HISQ} \\
\hline
$Ma$ & $n$ & $z_0^{(1)}$ & $\rho^{(1)}$ \\
\hline
3.297 & 4 & 0.024  & 0.082    \\
2.66 & 4 & 0.006  &   0.064   \\
1.91 & 4 & -0.007 &  0.051    \\
\hline
\hline
\end{tabular}
\caption{
Results for one-loop coefficients for the renormalisation 
of the lattice NRQCD-light temporal axial current for 
(from top to bottom) clover, asqtad and HISQ light quarks. 
Columns 1 and 2 given the bare lattice NRQCD mass and 
the stability parameter, $n$~\cite{Dowdall:2011wh}. 
For the NRQCD-clover results, $z_0^{(1)}$ is taken from~\cite{Znrqcdclover} 
where it is called $\rho_0$. For NRQCD-asqtad $z_0^{(1)}$ is 
obtained as $\tilde{\rho}_0-\zeta_{10}$ from~\cite{Znrqcdasqtad}. 
For NRQCD-HISQ $z_0^{(1)}$ is taken from~\cite{Dowdall:2013tga}. 
Values of $\rho^{(1)}$ make use of the appropriate $z_l^{(1)}$
as given in the text.
}
\label{tab:nrqcd}
\end{table}

An interesting question is whether this approach, in which 
the renormalisation constant for a mixed-action operator is 
defined in terms of renormalisation constants for the temporal 
vector current for the associated single-action operators,
also works for other actions in terms of giving a 
perturbative series for the remainder which is closer to 
1. Here we test this for the case of 
the heavy-light temporal axial current operator 
that combines an NRQCD~\cite{nrqcd} heavy quark with 
a light clover, asqtad or HISQ quark.  

Through order $\mathcal{O}(\Lambda/M)$ in an inverse heavy quark 
mass expansion, we define the renormalisation constant for 
the lattice NRQCD-light operator by~\cite{Dowdall:2013tga} 
\begin{equation}
A^4_{{\mathrm{cont\,QCD}}} = Z_{A^4,\mathrm{NRQCD}}(J^{(0)}+J^{(1)}).
\end{equation}
Here $J^{(0)}$ and $J^{(1)}$ are the leading and next-to-leading 
order operators in the $\Lambda/M$ expansion 
whose matrix elements between the vacuum and a $B$-meson are 
readily calculated in lattice QCD~\cite{Dowdall:2013tga}.
$Z_{A^4,\mathrm{NRQCD}}$ has been calculated through $\mathcal{O}(\alpha_s)$ 
for the combination of NRQCD heavy quarks with clover light quarks~\cite{Znrqcdclover}, asqtad light quarks~\cite{Znrqcdasqtad} and HISQ light 
quarks~\cite{Znrqcdhisq, Dowdall:2013tga} 
(in all cases setting the light quark mass 
to zero). Writing
\begin{equation}
\label{eq:znrqcd}
Z_{A^4,\mathrm{NRQCD-light}} = 1 + z_0^{(1)}\alpha_s + \ldots
\end{equation}
gives the values for $z_0$ given in Table~\ref{tab:nrqcd} 
for a selection of NRQCD bare quark masses in lattice units. 

A test of the renormalisation 
method advocated by the Fermilab collaboration 
is then to compare the perturbative series for $\rho_{A^4}$ 
where 
\begin{equation}
\label{eq:rhonrqcd}
\rho^{A^4,\mathrm{NRQCD-light}} = \frac{Z_{A^4,\mathrm{NRQCD-light}}}{\sqrt{Z_{V^4,\mathrm{NRQCD-NRQCD}}Z_{V^4,\mathrm{light-light}}}},
\end{equation}
as the analogue of Eq.~(\ref{eq:zhl}). 
Here we take for $Z_{V^4,\mathrm{NRQCD-NRQCD}}$ the renormalisation factor 
for the NRQCD-NRQCD temporal current in scattering, and this 
is 1 for equal masses~\cite{Boyle:2000fi}. The light-light 
renormalisation factor for massless quarks 
can be written as
\begin{equation}
\label{eq:zlight}
Z_{V^4,\mathrm{light-light}} = 1 + z_l^{(1)}\alpha_s + \ldots 
\end{equation}
and then, if 
\begin{equation}
\rho^{A^4,\mathrm{NRQCD-light}} = 1 + \rho^{(1)}\alpha_s + \ldots
\end{equation}
then 
$\rho^{(1)}=z_0^{(1)}-z_l^{(1)}/2$. The success of the method can then be 
judged by comparing the smallness of $\rho^{(1)}$ with that of $z_0^{(1)}$. 

Table~\ref{tab:nrqcd} gives values for $\rho^{(1)}$ for a variety 
of light-quark actions. 
For the tadpole-improved clover action, with a clover 
coefficient $c_{sw}=1$, on a gluon field 
from a simple plaquette action, which is appropriate 
to the $z_0$ calculation given in~\cite{Znrqcdclover}, 
we use~\cite{Luscher:1996jn, Capitani:2000xi} 
\begin{equation}
z_l^{(1)} = -1.6261 -  u_0^{(1)} 
\end{equation}
where $u_0^{(1)}$ is the one-loop coefficient of the 
tadpole parameter $u_0$ by which the gluon fields are 
divided. This division removes large and universal 
tadpole effects~\cite{Lepage:1992xa}. If the value 
of $u_0$ is taken as the fourth root of the 
average plaquette, then $u_0^{(1)} = -\pi/3$.  
This gives $z_l^{(1)}=-0.579$. 
For asqtad quarks, again using tadpole-improvement 
with $u_0$ set by the mean plaquette as appropriate 
to the $z_0$ values, and 
a Symanzik-improved gauge action, $z_l^{(1)}=0.411$~\cite{Kim:2010fj}. 
For HISQ quarks on Symanzik-improved gluon 
fields, $z_l^{(1)}=-0.1164(3)$ (Section~\ref{subsec:Zsf}).

We see from Table~\ref{tab:nrqcd} that 
the perturbative expansion of $\rho$ looks much 
better than that of $z_0$, as judged by the 
one-loop coefficients, for the NRQCD-clover and 
NRQCD-asqtad results. The effect of using 
eq.~(\ref{eq:rhonrqcd}) is to remove a constant 
factor coming from the light quark action 
in those cases (nothing is required for the NRQCD action). 
The remaining coefficients are generally then 
of approximately the same magnitude as the 
value quoted in Section~\ref{subsec:rho} for the 
massless clover-asqtad case, i.e. around 
0.05. 

For the NRQCD-HISQ case, the coefficient 
$\rho^{(1)}$ is 
actually larger in magnitude than 
$z_0^{(1)}$ and so, although $\rho^{(1)}$ 
is not large, it makes no sense to 
apply eq.~(\ref{eq:rhonrqcd}). The HISQ action 
needs so little renormalisation that 
correcting for a renormalisation issue 
that is not there is counterproductive. This 
is why in the current state-of-the-art determination 
using NRQCD 
of the $B$ and $B_s$ meson decay constants~\cite{Dowdall:2013tga}
and corresponding vector meson results~\cite{Colquhoun:2015oha} 
we simply applied the formula of eq.~(\ref{eq:znrqcd}) 
(along with additional current corrections that 
appear at $\alpha_s\Lambda/m_b$). 

\section{Z factors for the Future}
\label{subsec:Zsf}

\begin{figure}
\centering
\includegraphics[width=0.45\textwidth]{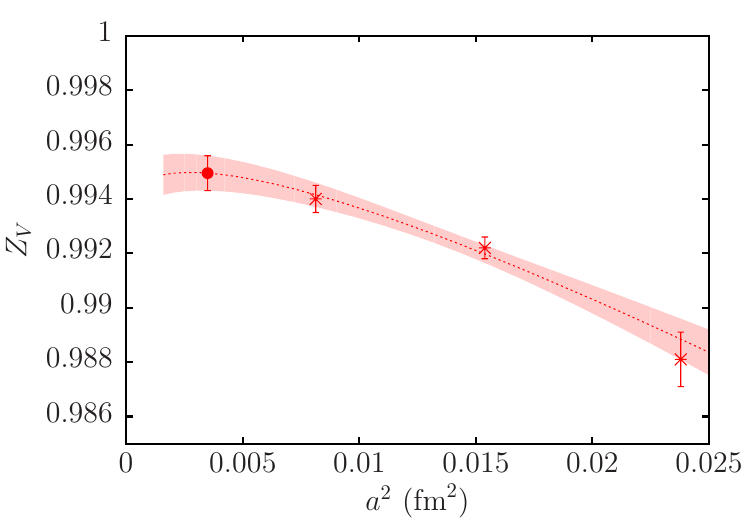}
\caption{ The renormalisation factor, $Z_V$, for the local 
temporal vector current between HISQ quark fields, plotted 
against the square of the lattice spacing. Results from 
Table~\ref{tab:results} are plotted as red bursts; the 
red band gives the fit to a perturbative expansion with 
discretisation effects described in the text. The red filled 
circle gives the extrapolated result on superfine lattices 
with $a$ = 0.059 fm.   
}
\label{fig:ZHH}
\end{figure}

Given that the $Z$ factors for the vector current can be determined very 
precisely by the nonperturbative method discussed in Section~\ref{subsec:V0} 
it is worth asking: a) how well can they be matched to perturbative 
expectations? and b) how well can we extrapolate the results to 
finer lattices? 
The reason for asking question b) is that the determination of $Z_V$ for 
the local HISQ temporal vector current using a 3-point function with 
clover spectator quark is numerically expensive. A numerically faster method is to 
use the RI-SMOM scheme~\cite{Sturm:2009kb}, being adapted 
for the HISQ action~\cite{Lytle:2015btr}, but here we investigate 
extrapolation as a way to remove the need for 
additional calculations.  
Since $Z_V$, as a perturbative expansion 
in $\alpha_s$, changes only slowly with lattice 
spacing it should be possible to extrapolate results to finer 
lattices without large uncertainties. Such an extrapolation, however, 
must include terms to allow for nonperturbative discretisation effects 
that will be present. 

To test this we fit the H-H $Z_V$ results from Table~\ref{tab:results} 
to the following form: 
\begin{equation}
\label{eq:pert}
Z_V(a,\alpha_s) = \sum_{i=0}^{n_i} \left[c_i+d_i(\frac{a\Lambda}{\pi})^2 + f_i(\frac{a\Lambda}{\pi})^4\right]\alpha_s^i 
\end{equation}
where $\alpha_s$ is taken in the $\overline{\mathrm{MS}}$ scheme at a scale of $2/a$, although 
using $1/a$ or $3/a$ makes little difference. 
$c_0$ is taken as 1.0 and $c_1$ = -0.1164(3) from lattice 
QCD perturbation theory~\cite{Trottier}. 
The other $c_i$, and the $d_i$ and $f_i$, are given 
priors of 0.0(1.0). 
We take $\Lambda$ = 0.5 GeV and a discretisation effect dependent on
$a\Lambda/\pi$ suitable for an ultraviolet quantity such as $Z$, 
since the momentum cut-off 
on the lattice is $\pi/a$. Using $n_i = 5$ gives 
the fit curve plotted in Figure~\ref{fig:ZHH}; increasing $n_i$ beyond this 
makes no difference, and decreasing $n_i$ to 3 also has little impact.
The fit has a $\chi^2/\mathrm{dof}$ of 0.8 and favours a positive 
 and fairly sizeable (although quite uncertain) 
coefficient at $\alpha_s^2$ of 0.59(16). 

This enables us to predict the value of $Z_V$ on superfine 
lattices (with $a =$ 0.059fm) with an accuracy of 0.06\% 
as 0.9950(6). The dominant uncertainty (0.06\%) comes from 
the statistical uncertainties on the $Z$ factors on coarser 
lattices, with 0.01\% coming from the $a^2$ extrapolation 
and 0.02\% from the perturbative series.   
We have checked that the extrapolated superfine result 
is not affected significantly (less than $1\sigma$) by 
missing out the value on the coarsest lattices from the fit. 
A further check of the result and its uncertainty 
comes from fitting the results on the coarsest two 
lattices to predict a value for the fine lattice. This 
gives 0.9948(10), 
in good agreement (within $1\sigma$) of the actual value 
we have calculated there of 0.9940(5) (see Table~\ref{tab:results}). 
Using our fit to the full set of results we can also obtain 
a value for $Z_V$
for ultrafine lattices (with $a =$ 0.044 fm) 
of 0.9949(7). 

These uncertainties are small enough to mean that the $Z$ 
factors will not cause a dominant uncertainty in the 
calculation, for example, of the hadronic vacuum polarisation 
contribution to the anomalous magnetic moment of 
the muon on these finer lattices~\cite{Chakraborty:2016mwy}. 
Note that these results are for a current composed of $s$ quarks. 
We have not studied the dependence on quark mass of the $Z$ 
factors; from perturbation theory it should be small, with a 
leading term of $\alpha_s(ma)^2$. 

\bibliography{zpaper}

\end{document}